%% file: huukerr.tex
\documentclass[
	aps,
	prd, 
	twocolumn, 
	a4paper, 
	10pt, 
	amsmath,
	amssymb, 
	preprintnumbers, 
	tightenlines, 
	notitlepage,
	floatfix,
	nofootinbib, 
	longbibliography
	]{revtex4-1}\pdfoutput=1
\usepackage[english]{babel}
\usepackage{bbm}
\usepackage{graphicx}
\usepackage{hyperref}
\usepackage{wasysym}
\usepackage{amsthm}
\usepackage{dcolumn}
\usepackage[normalem]{ulem}
\usepackage{color}
	\newcolumntype{.}{D{.}{.}{13}}
	\newcolumntype{d}[1]{D{.}{.}{#1}}
	
\usepackage[all,cmtip]{xy}

\graphicspath{{pics/}}

\allowdisplaybreaks[1]

\newcommand{\TabH}[1]{\multicolumn{1}{c}{#1}}

\newcommand{\abs}[1]{\lvert#1\rvert}			
\newcommand{\avg}[1]{\langle #1\rangle }		

\newcommand{\cbB}[1]{\Big\{#1\Big\} }			
\newcommand{\hh}[1]{\left(#1\right) }			
\newcommand{\nh}[1]{(#1) }						
\newcommand{\bh}[1]{\bigl(#1\bigr) }			
\newcommand{\Bh}[1]{\Bigr(#1\Bigr) }			

\newcommand{\id}[1]{\operatorname{d}\!#1}												
\renewcommand{\d}[2]{\frac{\operatorname{d}\!#1}{\operatorname{d}\!#2}}					
\newcommand{\CD}[1]{\nabla_{#1}}

\newcommand{\ii}{i}								
\newcommand{\ee}{e}								

\newcommand{\pt}{\tau}							
\newcommand{\mt}{\lambda}						

\newcommand{\nE}{\mathcal{E}}								
\newcommand{\nL}{\mathcal{L}}							

\newcommand{\rmin}{r_\mathrm{min}}
\newcommand{\rmax}{r_\mathrm{max}}
\newcommand{\zmax}{z_\mathrm{max}}

\newcommand{\mr}{\mu}

\newcommand{\ptrp}{\mathcal{T}_r}
\newcommand{\ctrp}{T_r}
\newcommand{\mtrp}{\Lambda_r}

\DeclareMathOperator{\bigO}{\mathcal{O}}					

\newcommand{\tet}[2]{e_{#1}^{#2}}				

\newcommand{\R}[3][]{\,{_{#2}R_{#3}^{\mathrm{#1}}}}		
\newcommand{\Rb}[3][]{\,{_{#2}\bar{R}_{#3}^{\mathrm{#1}}}}		
\newcommand{\SWSH}[3][]{\,{_{#2}S_{#3}^{\mathrm{#1}}}}	
\newcommand{\Y}[3][]{\,{_{#2}Y_{#3}^{\mathrm{#1}}}}		
\newcommand{\T}[3][]{\,{_{#2}T_{#3}^{\mathrm{#1}}}}		

\newcommand{\Sb}[4]{({_{#1}b_{#2}})^{#3}_{#4}}			
\newcommand{\YA}[4]{{^{#1#2}\!\!\mathcal{A}^{#3}_{#4}}}			
\newcommand{\YB}[4]{{^{#1#2}\!\mathcal{B}^{#3}_{#4}}}			

\newcommand{\TEV}[2]{\,{_{#1}\lambdabar_{#2}}}			
\newcommand{\SEV}[2]{\,{_{#1}A_{#2}}}					

\newcommand{\A}[3][]{{_{#2}A_{#3}^{\mathrm{#1}}}}		
\newcommand{\B}[3][]{{_{#2}B_{#3}^{\mathrm{#1}}}}		

\newcommand{\Ainf}[3][]{{_{#2}\alpha_{#3}^{\mathrm{#1}}}}	
\newcommand{\Binf}[3][]{{_{#2}\beta_{#3}^{\mathrm{#1}}}}	

\newcommand{\Ab}[3][]{{_{#2}\bar{A}_{#3}^{\mathrm{#1}}}}		
\newcommand{\Bb}[3][]{{_{#2}\bar{B}_{#3}^{\mathrm{#1}}}}		

\newcommand{\Ainfb}[3][]{{_{#2}\bar{\alpha}_{#3}^{\mathrm{#1}}}}	
\newcommand{\Binfb}[3][]{{_{#2}\bar{\beta}_{#3}^{\mathrm{#1}}}}	

\newcommand{\ph}[2]{(#1)_{#2}}			

\newcommand{\pI}{+}						
\newcommand{\uI}{\not{+}}				
\newcommand{\pH}{-}						
\newcommand{\uH}{\not{-}}				
\newcommand{\pIH}{\pm}					
\newcommand{\uIH}{\not{\pm}}			
\newcommand{\uHI}{\not{\mp}}			

\newcommand{\F}{{_{2}F_1}}				
\newcommand{\lP}[3]{P_{#1}^{#2}(#3)}	
\newcommand{\lQ}[3]{Q_{#1}^{#2}(#3)}	

\newcommand{\MROP}[1]{\hat{\mathcal{H}}_{#1}}	

\newcommand{\MC}{\mathcal{C}}					

\newcommand{\SWL}[1]{\bar\eth_{#1}}				
\newcommand{\TP}{\epsilon}						

\newcommand{\lmax}{l_\mathrm{max}}				
\newcommand{\llmax}{l_{1,\mathrm{max}}}			
\newcommand{\kmax}{k_\mathrm{max}}				
\newcommand{\Lmax}{L_\mathrm{max}}				
\newcommand{\Lmin}{L_\mathrm{min}}				

\newcommand{\ret}{\mathrm{Ret}}

\newcommand{\reg}{\mathrm{R}}
\newcommand{\sing}{\mathrm{S}}

\newcommand{\diss}{\mathrm{diss}}
\newcommand{\cons}{\mathrm{cons}}

\newcommand{\Lor}{\mathrm{Lor}}
\newcommand{\Rad}{\mathrm{Rad}}

\newcommand{\RP}[1]{\mathcal{E}_{#1}}

\newcommand{\pISO}{p_\mathrm{ISO}}
\begin{document}

\title{Metric perturbations produced by eccentric equatorial orbits around a Kerr black hole}

\author{Maarten \surname{van de Meent}}
\email{M.vandeMeent@soton.ac.uk}
\author{Abhay G. \surname{Shah}}
\email{a.g.shah@soton.ac.uk}
\affiliation{Mathematical Sciences, University of Southampton, Southampton, SO17 1BJ, United Kingdom}

\date{\today}
\begin{abstract}
We present the first numerical calculation of the (local) metric perturbation produced by a small compact object moving on an eccentric equatorial geodesic around a Kerr black hole, accurate to first order in the mass ratio. The procedure starts by first solving the Teukolsky equation to obtain the Weyl scalar $\psi_4$ using semi-analytical methods. The metric perturbation is then reconstructed from $\psi_4$ in an (outgoing) radiation gauge, adding the appropriate non-radiative contributions arising from the shifts in mass and angular momentum of the spacetime.

As a demonstration we calculate the generalized redshift $U$ as a function of the orbital frequencies $\Omega_r$ and $\Omega_\phi$ to linear order in the mass ratio, a gauge invariant measure of the conservative corrections to the orbit due to self-interactions. In Schwarzschild, the results surpass the existing result in the literature in accuracy, and we find new estimates for some of the unknown 4PN and 5PN terms in the post-Newtonian expansion of $U$. In Kerr, we provide completely novel values of $U$ for eccentric equatorial orbits. Calculation of the full self-force will appear in a forthcoming paper.
\end{abstract} 

\maketitle
\setlength{\parindent}{0pt} 
\setlength{\parskip}{6pt}

\section{Introduction}


With Advanced LIGO scheduled to start data collection this year \cite{advLIGO}, the detection of gravitational waves seems imminent. One of the main expected sources of gravitational waves are binary systems of compact objects in close orbits. Detection of gravitational waves from these systems is dependent on our ability to model them accurately, requiring us to solve the two-body problem including general relativistic effects. Unlike its Newtonian counterpart, the general relativistic two-body problem admits no analytic solutions and requires some sort of approximation scheme to solve. There are currently three main approaches, which are largely complementary.

Numerical relativity (NR) works by discretizing spacetime and numerically solving the nonlinear field equations of general relativity on a grid. Spectacular progress has been made in this approach over the last few decades (see \cite{Choptuik:2015mma} and the references therein). However, discretization of spacetime means that this approach has limited use in systems with vastly differing length and time scales.

Post-Newtonian (PN) theory (effectively) expands the equations of motion in the binary separation, obtaining corrections to Newtonian theory order-by-order (see \cite{Blanchet:2013haa} and the references therein). This provides extremely good models for the motion at large separations. But as one enters the strong gravity regime, convergence of the post-Newtonian expansion is poor. 

Black hole perturbation theory (used in this paper) approximates the motion of the binary by expanding in the mass-ratio of the components, starting from test particle (geodesic) motion in the geometry generated by the largest mass\cite{Poisson:2011nh,Barack:2009ux,Pound:2015}. This is most naturally applied to systems with a very small mass ratio, such as extreme mass ratio inspirals (EMRIs) formed by a stellar mass compact object spiralling towards a (super)massive black hole, expected to occur regularly in galactic cores. The gravitational waves produced by an EMRI fall outside the sensitivity range of ground based interferometers (LIGO/Virgo/KAGRA), but are expected to be one of the prime sources for the space-based gravitational wave observatory eLISA, slated for launch in the 2030s \cite{eLISA}.

In the meantime, black hole perturbation theory can be used to validate and inform NR and PN techniques by comparing results in regions of the compact binary parameter space where its range of applicability overlaps with the other methods. To compare results between different approximation methods requires studying quantities that are insensitive to the approximation used. In particular, since different approaches tend to work in different gauges, we need the quantities to be gauge invariant.

In addition, results can be used to tune the effective-one-body (EOB) formalism developed by \cite{Buonanno:1998gg}. This formalism takes results from PN theory, black hole perturbation theory and numerical relativity to reduce the two-body problem to an effective analytical model of a single point particle moving in an effective metric \cite{Damour:1999cr} similar to that of the Newtonian reduced mass problem. Advances have been made with the latest being the evolution of a non-spinning binary-neutron-star by \cite{Bernuzzi:2014owa,Hotokezaka:2015xka} whose results agree to those of NR within numerical accuracy. The free parameters in this theory are fixed by comparing gauge-invariants of physical interest from the above three approaches. 

First such invariant to be calculated was the ``redshift'' invariant \cite{Detweiler:2008ft}, the linear-in-mass-ratio change in the time-component of the 4-velocity of the compact object in circular orbit around a Schwarzschild black hole which was then used to calibrate the binding energy and angular momentum of a binary in the EOB model \cite{LeTiec:2011ab}. Later on, many other invariants have been calculated for circular orbits, including corrections to the circular limit of the periastron advance\cite{Barack:2010ny}, change in innermost-stable-circular orbit \cite{Barack:2009ey,Isoyama:2014mja}, spin precession \cite{Dolan:2013roa}, quadrupolar-tidal \cite{Dolan:2014pja}, and octupolar-tidal invariants  \cite{Nolan:2015vpa}. PN expansions of the latter three have been studied to very high order \cite{Nolan:2015vpa,Shah:2015nva}, and were used to calibrate the EOB model further \cite{Bini:2014nfa,*Bini:2014ica,*Bini:2014zxa,*Bini:2015bla,*Bini:2015mza}.

Calculations of the redshift invariant for circular orbit in Schwarzschild spacetime have been generalized to eccentric orbits around a Schwarzschild black hole where they calculated the orbit-averaged value of the invariant, $U$, for given azimuthal and radial frequencies using a Lorenz gauge in the time-domain \cite{Barack:2011ed}, and recently, a successful comparison was performed for this orbit-averaged quantity between the self-force and PN theories using frequency domain methods \cite{Akcay:2015pza}.

All these calculations for orbits around a non-rotating (Schwarzschild) black hole have been performed either in a Lorenz gauge or in the Regge-Wheeler-Zerilli gauge (or both). In both gauges, the linearized Einstein equation is separable. In the frequency-domain, they may be solved mode-by-mode by solving second order ordinary differential equations.

A major obstacle in generalizing these calculations to rotating (Kerr) black holes has been that there are no known gauges for which the linearized Einstein equation in Kerr spacetime is separable. In recent years, an alternative approach has been put forward that exploits the fact that the Teukolsky equation for the Weyl scalars $\psi_0$ and $\psi_4$ \emph{are} separable and can be solved efficiently. Using the formalism developed by Chrzanowski \cite{Chrzanowski:1975wv}, Cohen and Kegeles \cite{Cohen:1974cm,Kegeles:1979an}, and Wald \cite{Wald:1978vm} (CCK or CCKW), one can reconstruct the metric perturbation in a radiation gauge from the Weyl scalars.

This approach has been implemented to obtain the metric perturbation produced by a particle moving on a circular orbit around Schwarzschild \cite{Keidl:2010pm,Shah:2010bi}, and Kerr black holes \cite{Shah:2012gu}. Using this technique high precision results for some of the invariants above have been obtained in the Schwarzschild case and employed to extract very high PN-expansion using a numerical fitting technique \cite{Shah:2013uya,Shah:2015nva,Johnson-McDaniel:2015vva}. In the Kerr case, the shift in the inner-most-stable circular equatorial orbit \cite{Isoyama:2014mja}, and the redshift invariant \cite{Shah:2012gu} have been calculated. Work on the PN-expansion of the latter is in progress \cite{Shah:???}.

Despite the success of \cite{Keidl:2010pm,Shah:2010bi,Shah:2012gu}, it was unclear that the self-force corrections to the motion that they calculated were well-defined due to radiation gauge metric perturbations being irregular in a neighbourhood of the particle worldline. The use of radiation gauge metric perturbations to obtain self-force corrections to the motion of a particle was put on a firm formal footing in \cite{Pound:2013faa}. In particular, they showed how the mode-sum regularisation formula for the self-force needs to be modified to use radiation gauge metric perturbations as its input.

The goal of this paper is to implement the formalism set out in \cite{Pound:2013faa} for eccentric (equatorial) orbits to obtain---for the first time---the metric perturbation produced by a particle moving on an eccentric orbit around a Kerr black hole. A problem with extending this approach to eccentric orbits is that the CCK procedure for obtaining the metric perturbation from the Weyl scalars is only well-defined for vacuum spacetimes. However, when solving the Teukolsky equation sourced by a particle on an eccentric orbit in the frequency domain, the particle point source gets smeared out of the entire region between the periapsis $\rmin$ and apapsis $\rmax$ of the orbit. We overcome this problem utilizing the method of extended homogeneous solutions \cite{Barack:2008ms} to avoid dealing with non-vacuum solutions at any stage of the calculation.

Our procedure will be as follows. We obtain highly accurate solutions to the Teukolsky equation using the semi-analytical Mano-Suzuki-Takasugi (MST) formalism. We then algebraically invert the fourth order differential equation for the intermediate Hertz potential, from which we obtain the (singular) metric perturbation using the CCK procedure. The regular part of the metric perturbation is obtained using mode-sum regularization. The missing pieces of the metric perturbation due to perturbations of the mass and angular momentum of the background spacetime, which cannot be recovered using the CCK procedure, are obtained using the result from \cite{Merlin:2015}.

Using the obtained (regular part of the) metric perturbation we calculate the self-force correction to generalized redshift invariant $U$ for eccentric orbits as a function of the orbital frequencies. In Schwarzschild, our results match to the previously published results of \cite{Barack:2011ed} and \cite{Akcay:2015pza} to all given digits, surpassing them in accuracy and computational efficiency. By fitting to a large dataset of orbits, we are able to recover all known coefficients of the PN expansion of the generalized redshift in Schwarzschild and obtain estimates for some of the unknown 4PN and 5PN terms.  

In Kerr, our result matches the previously published in circular limit \cite{Shah:2012gu}. Our completely novel results for the generalized redshift for eccentric equatorial orbits around a Kerr black hole, pass all the consistency checks that we can perform. In particular, we numerically  match the analytically known leading regularization parameters. Besides verifying our numerical method, this provides the first every test of their analytical calculation for eccentric orbits in Kerr.

The paper is organized as follows. In section \ref{sec:prelim} we review some of the background need to setup our calculation, introducing the notation and conventions we will need. In particular, we will define the generalized redshift invariant that we will be calculating. Section \ref{sec:method} then describes the method that we will use to calculate the metric perturbations generated by a particle orbiting a Kerr black hole on an equatorial eccentric orbit. Details of the numerical implementation of this method are given in section \ref{sec:implementation}. In section \ref{sec:results} we numerically calculate the generalized redshift invariant $U$. We first compare the results from our code to values in the existing literature in limits of eccentric orbits around a Schwarzschild black hole and circular equatorial orbits around a Kerr black hole. We then give the first ever numerical values of $U$ for eccentric equatorial orbits around a spinning black hole. We conclude in section \ref{sec:discussion} with a discussion of the limitations of our code, and how it can be extended to calculate different quantities derived from the metric perturbations including the full (first order) self-force correction to the motion.

\section{Preliminaries}\label{sec:prelim}
\subsection{Some conventions and notation}
In this paper, we use geometric units $c=G=1$. In addition, we will usually set the mass of the central black hole, $M$ to $1$ as well.

Whenever a sum is written without explicit bounds on the indexed summed over, it is assumed to be summed over its full natural range, i.e. all (integer) values of the index for which the summand is well-defined and non-zero. For example, sums over $\ell$ will typically range from $\abs{s}$ to $\infty$, sums over $m$ will range from $-\ell$ to $\ell$, etc.

When in this paper we refer to (spin-weighted) spherical/spheroidal harmonics, we generally mean the ``polar'' part of the function, i.e. the part that depends on the polar coordinate $\theta$ (or $z$, see below). These functions are normalized such that 
\begin{equation}
\int_{-1}^{1}\id{z} \Y{s}{lm}(z)^2 = 1.
\end{equation}
The usual spherical harmonics are therefore
\begin{equation}
\Y{}{lm}(\theta,\phi)=\frac{\Y{0}{lm}(\cos\theta)\ee^{\ii m\phi}}{\sqrt{2\pi}}.
\end{equation}
We use overbars to denote complex conjugation, i.e. $\bar{x}$ is the complex conjugate of $x$.

%
In this paper, we write the metric generated by a rotating black hole with mass $M=1$ and spin $a$ as,
\begin{equation}
\begin{split}
\label{eq:kerr}
g_{\mu\nu} = 
-\bh{1 - \frac{2r}{\Sigma}}\id{t}^2 
+ \frac{\Sigma}{\Delta} \id{r}^2
+ \frac{\Sigma}{1-z^2} \id{z}^2
\\
+ \frac{1-z^2}{\Sigma} \bh{2a^2 r (1-z^2)+(a^2+r^2)\Sigma}\id\phi^2
\\
- \frac{4ar(1-z^2)}{\Sigma}\id{t}\id\phi,
\end{split}
\end{equation}
with
\begin{align}
\Delta &= r(r-2) + a^2,\\
\Sigma &= r^2 + a^2 z^2.
\end{align}
Here $t$, $r$, and $\phi$ are the usual Boyer-Lindquist coordinates, while $z$ is related to the usual  Boyer-Lindquist coordinate $\theta$ by $z=\cos\theta$. In particular, the equator of our black hole spacetime is given by $z=0$. The two roots of $\Delta$ indicating the location of the inner and outer horizon are denoted
\begin{equation}
 r_\pm =  1 \pm \sqrt{1-a^2}.
\end{equation}

Much of the techniques used in this paper rely on the Newman-Penrose formalism. We use the following null tetrad (the Kinnersley  tetrad in our modified Boyer-Lindquist coordinates),
\begin{alignat}{3}
\tet{1}{\mu} &= l^\mu &&= \frac{1}{\Delta}(r^2+a^2,\Delta,0,a),\\
\tet{2}{\mu} &= n^\mu &&= \frac{1}{2\Sigma}(r^2+a^2,-\Delta,0,a),\\
\tet{3}{\mu} &= m^\mu &&= -\frac{\bar\rho \sqrt{1-z^2}}{\sqrt{2}}(\ii a,0,-1,\frac{\ii}{1-z^2}),\\
\tet{4}{\mu} &= \bar{m}^\mu &&= \frac{\rho \sqrt{1-z^2}}{\sqrt{2}}(\ii a,0,1,\frac{\ii}{1-z^2}),
\end{alignat}
with
\begin{equation}
\rho = \frac{-1}{r-\ii a z}.
\end{equation}
We will use Greek letters to represent spacetime indices and Latin letters for tetrad indices. In particular, 
\begin{equation}
g_{ab} = \tet{a}{\mu}g_{\mu\nu}\tet{b}{\nu} =\begin{pmatrix}
0	&	-1	&	0	&	0\\
-1	&	0	&	0	&	0\\
0	&	0	&	0	&	1\\
0	&	0	&	1	&	0\\
\end{pmatrix},
\end{equation}
is the metric in tetrad indices, and will be used to raise and lower tetrad indices.
\subsection{Two-body problem in black hole perturbation theory}
We are interested in the motion of a pair of gravitationally bound compact masses $m$ and $M$, in the limit that $m\ll M$.\footnote{With some abuse of notation we will use $m$ and $M$ both as labels for the two objects and as the value of their masses.} In this limit, the gravitational field produced by the smaller object $m$ can be treated as a perturbation to the black hole geometry generated by the larger object $M$. If we place ourselves in a frame in which $M$ is at rest, we can study the motion of $m$ as a perturbative series in the mass ratio $\mr=m/M\ll 1$.

Both masses are assumed to be compact enough that they can be treated as black holes. Moreover we assume both masses to have no charge. We allow the larger mass to have non-zero spin $a=J/M$. The background geometry is therefore described by the Kerr metric, which in addition is assumed to be sub-extremal, i.e. $a/M < 1$. We further assume the smaller object to have zero spin.

\subsection{Geodesics in Kerr spacetime}
In the limit $\mr\to 0$, the smaller object will follow a geodesic of the Kerr spacetime generated by the larger object $M$. The geodesic equations in Kerr spacetime have the following form:
\begin{align}
\Bh{\d{r}{\pt}}^2 &=\frac{ R(r)}{\Sigma(r,z)^2}, \\
\Bh{\d{z}{\pt}}^2 &= \frac{Z(z)}{\Sigma(r,z)^2},  \\
\d{\phi}{\pt} &=\frac{\Phi_r(r)+\Phi_z(z)}{\Sigma(r,z)},  \\
\d{t}{\pt} &= \frac{\mathfrak{T}_r(r)+\mathfrak{T}_z(z)}{\Sigma(r,z)},  
\end{align}
where $R$, $Z$, $\Phi_r$, $\Phi_z$, $\mathfrak{T}_r$, and $\mathfrak{T}_z$ are known functions of $r$ and $z$ (see e.g. \cite{Fujita:2009bp}).

As first shown by Carter \cite{Carter:1968rr}, this set of equations is separable. This separation can be achieved easily, by changing to a convenient time variable $\mt$ to parametrize the orbit,
\begin{equation}
 \d{\pt}{\mt}=\Sigma(r,z) =  r^2+a^2z^2.
\end{equation}
The time variable $\mt$ is commonly referred to as ``Mino time''. This parametrization allows for an analytic solution of geodesic equations in terms of elliptic functions \cite{Fujita:2009bp}.

The solutions in \cite{Fujita:2009bp} are most naturally expressed in terms of the periapsis $\rmin$, the apapsis $\rmax$, and the maximum value of the $z$ coordinate $\zmax$. In this paper we are concerned only with equatorial orbits. Consequently, we will set $\zmax=0$. Instead of the parameters $(\rmin, \rmax)$ we use the semilatus rectum $p$ and eccentricity $e$, defined by
\begin{align}
\rmin &= \frac{p}{1+e},\\
\rmax &= \frac{p}{1-e},
\end{align}
to indentify orbits.

To compare orbits between different spacetimes we need a gauge invariant way of identifying orbits. A useful set of invariants is given by the orbital frequencies \citep{Barack:2011ed},
\begin{align}
\Omega_r 	&= \frac{
	2\pi
	}{
	\ctrp
	}, \\
\Omega_\phi &= \frac{
	\avg{\d{\phi}{\pt}}
	}{
	\avg{\d{t}{\pt}}
	},
\end{align} 
where $\ctrp$ is Boyer-Lindquist coordinate time between two passes through periapsis, and the angular brackets $\avg{\cdot}$ denote averaging with respect to proper time,
\begin{equation}
\avg{F} \equiv \lim_{\mathcal{T}\to\infty} \frac{1}{2\mathcal{T}}\int_{-\mathcal{T}}^{\mathcal{T}} F(\pt) \id\pt.
\end{equation}
For future reference we note that for periodic quantities this definition reduces to,
\begin{align}
\avg{F} &=\frac{1}{\ptrp}\int_0^{\ptrp} F(\pt) \id\pt\\
&=\frac{1}{\ptrp}\int_0^{\mtrp} F(\mt)\Sigma \id\mt,
\end{align}
where $\ptrp$ is the proper time between two passes through periapsis, and $\mtrp$ is the corresponding amount of Mino time.

The pair $(\Omega_r,\Omega_\phi)$ provides a gauge independent characterization of eccentric equatorial orbits. This characterization is almost (but not quite) unique. In a small region near the transition to plunging orbits, all orbits come in isofrequency pairs of distinct orbits with identical frequencies $(\Omega_r,\Omega_\phi)$ \cite{Warburton:2013yj}.

\subsection{Self-force corrected motion}\label{sec:GSF}
We now consider the situation that $0<m<<M(=1)$. In that limit, we can study the effect of the small mass $m$ on the gravitational field as a perturbation to the black hole geometry generated by the larger mass $M$, using the mass-ratio $\mr=m/M$ as an expansion parameter. The corrections to the motion of the smaller object due to self-interactions can thus also be studied order-by-order in $\mr$. Detweiler and Whiting \cite{Detweiler:2002mi} showed that, to first-order in $\mr$, the mass $m$ will follow a geodesic in the effective spacetime
\begin{equation}
\tilde{g}_{\mu\nu} = g_{\mu\nu} + h_{\mu\nu}^\reg,
\end{equation}
where $g_{\mu\nu}$ is the background Kerr geometry generated by the larger mass $M$, and $h_{\mu\nu}^\reg$ is a certain smooth piece of the retarded metric perturbation generated by the smaller mass $m$.

The regular piece of the metric perturbation, $h_{\mu\nu}^\reg$,  can be further split in a dissipative piece $h_{\mu\nu}^{\reg,\diss}$, and a conservative piece $h_{\mu\nu}^{\reg,\cons}$. By setting appropriate boundary conditions on the retarded metric perturbation, the effects of the dissipative and conservative piece can be studied independently. The dissipative piece encodes all effects on the motion of $m$ due to energy and angular momentum being carried away by gravitational waves causing the orbit to slowly decay and spiral towards the larger mass $M$. The conservative piece then encodes all other effects like changes in the orbital periods.

Because the motion in the ``conservative''  effective spacetime,
\begin{equation}
\tilde{g}_{\mu\nu}^\cons = g_{\mu\nu} + h_{\mu\nu}^{\reg,\cons},
\end{equation}
is free from dissipation, there exist bound eccentric equatorial orbits in the effective spacetime with well-defined orbital frequencies
\begin{align}
\tilde\Omega_r 	&= \frac{
	2\pi
	}{
	\tilde\ctrp
	}, \\
\tilde\Omega_\phi &= \frac{
	\avg{\d{\tilde\phi}{\tilde\pt}}
	}{
	\avg{\d{\tilde t}{\tilde\pt}}
	},
\end{align}
where the tildes denote quantities calculated in the effective spacetime.

\subsection{Generalized redshift invariant}
To compare the conservative effect of the self-interaction between various computational approaches one needs a gauge invariant measure of this effect. In this paper we will calculate one such measure, the so-called \emph{(generalized) redshift}. This was first introduced for circular orbits by Detweiler \cite{Detweiler:2008ft}, and later generalized to eccentric orbits by Barack and Sago \cite{Barack:2011ed,Akcay:2015pza}. Recent work by Le Tiec \cite{LeTiec:2015cxa} has shown that the generalized redshift plays a key role in the `first law of mechanics' for compact binaries, where it acts as a conjugate variable to the particle mass.  In its generalized form the redshift invariant is defined as,
\begin{equation}
U\equiv \avg{\d{t}{\pt}}.
\end{equation}
For circular orbits (and a helically symmetric choice of gauge) $\d{t}{\pt}$ is independent of $\pt$ and this definition reduces Detweiler's original definition 
\begin{equation}
U = \d{t}{\pt}.
\end{equation}
For eccentric orbits we have,
\begin{equation}
U= \frac{1}{\ptrp}\int_0^{\ptrp} \d{t}{\pt}\id\pt = \frac{\ctrp}{\ptrp}.
\end{equation}
The redshift invariant $U$ is invariant under gauge transformations that are asymptotically flat and preserve the periodic nature (or helical symmetry in the case of circular orbits) of the spacetime \cite{Barack:2011ed}. Consequently, $U$ viewed as a function of the orbital frequencies  $(\Omega_r,\Omega_\phi)$ provides a gauge invariant measure of the conservative motion. We can thus compare $U$ in the background spacetime $g_{\mu\nu}$ and the conservative effective spacetime $\tilde{g}_{\mu\nu}^\cons$ (at fixed orbital frequency),
\begin{equation}
\Delta U(\Omega_r,\Omega_\phi) \equiv \frac{\tilde{U}(\Omega_r,\Omega_\phi) - U(\Omega_r,\Omega_\phi)}{\mr}.
\end{equation}
Following \citep{Akcay:2015pza} we obtain $U$ in terms of quantities that we can calculate directly for perturbation theory,
\begin{align}
U &=\frac{1}{\mr}\hh{ \frac{\ctrp}{\tilde\ptrp}-\frac{\ctrp}{\ptrp}}
\\
&= -\frac{\ctrp}{\mr\ptrp^2} \hh{\tilde\ptrp-\ptrp}\id\pt +\bigO(\mu)
\\
&= -\frac{\ctrp}{\mr\ptrp^2}\int_0^{\ptrp} (\d{\tilde\pt}{\pt}-1)\id\pt +\bigO(\mu).
\end{align}
From the normalization conditions for the 4-velocity $u^\mu=\d{x^\mu}{\pt}$,
\begin{align}
g_{\mu\nu}u^\mu u^\nu = \tilde{g}_{\mu\nu}^\cons \tilde{u}^\mu \tilde{u}^\nu = -1,
\end{align}
one readily obtains
\begin{equation}
\d{\tilde\pt}{\pt} = 1 - \frac{1}{2} h_{\mu\nu}^{R,\cons}u^\mu u^\nu +\bigO(\mr^2),
\end{equation}
and consequently
\begin{equation}
\Delta U = \frac{\ctrp}{2\mr\ptrp} \avg{h_{uu}^R},
\end{equation}
where $h_{uu}^R \equiv h_{\mu\nu}^{R}u^\mu u^\nu= h_{\mu\nu}^{R,\cons}u^\mu u^\nu$.

\subsection{Radiation gauge}
In this paper, we will obtain the metric perturbation $h_{\mu\nu}$ in the so-called \emph{outgoing radiation gauge} or ORG. This gauge is defined by the gauge conditions
\begin{align}
\tet{2}{\mu}h_{\mu\nu} &= 0,\text{ and}\\
h_{\mu}^\mu 	&= 0.
\end{align}
This gauge is naturally asymptotically flat, and the gauge conditions respect the periodic nature of orbits in the background spacetime. As such the ORG is a natural gauge to calculate $\Delta U$.

A complicating factor of work in radiation gauges is that they are known \cite{Barack:2001ph,Ori:2002uv} to develop singularities in the presence of matter, and this singularity will generally extend away from the worldline. This issue was studied in great detail by Pound et al. \cite{Pound:2013faa}. They found that radiation gauges fall in one of four classes. In the first two classes, the gauge irregularity forms a string-like singularity extending along one of the principle null-directions, either towards infinity or the horizon. These are known as half-string gauges. In the so-called full-string gauge the singularity extends in both directions. Finally, one can stitch together two regular halves from the half-string gauges to form a gauge that is regular both at the horizon and infinity. The price paid is that this ``no-string'' gauge is discontinuous on a sphere containing the particle worldline.

In this paper, we obtain the local metric perturbation generated by a particle by approaching the particle's worldline either from the inside or outside in the half-string gauge that is regular on that side. The inside and outside limits are therefore obtained in different gauges. This will not cause problems as long as we remain aware that the limits of gauge dependent quantities may not agree.

\section{Method}\label{sec:method}
\subsection{Strategy}
The core idea of our strategy is to separate the metric perturbation in modes depending on just one coordinate variable, and find these functions mode-by-mode. Such `frequency domain' methods have previously been used to great success to solve the linearized Einstein equation in a Lorenz gauge for small objects orbiting a Schwarzschild black hole \cite{Barack:2008ms,Akcay:2010dx,Akcay:2013wfa}. Although this method has previously been applied to scalar waves on a Kerr background \cite{Warburton:2010eq,Warburton:2011hp,Warburton:2014bya}, it cannot be applied directly to find perturbations to the metric on a Kerr background, because the linearized Einstein equation in Kerr is not separable. Instead we have to work around this issue.
 
Teukolsky famously showed that the linear equations for scalar fields with arbitrary spin-weight on a Kerr background are separable \cite{Teukolsky:1973ha}. In particular, this holds for the equations for the Weyl scalars $\psi_0$ and $\psi_4$, which have spin-weight $+2$ and $-2$, respectively. In a 1973 theorem \cite{Wald:1973}, Wald showed that the perturbations to the Weyl scalars $\psi_0$ and $\psi_4$ individually contain almost all information about the metric perturbations (up to global perturbations of the mass and angular momentum). Cohen, Chrzanowski, and Kegeles (CCK) \cite{Cohen:1974cm,Chrzanowski:1975wv,Kegeles:1979an,Wald:1978vm} developed a procedure to reconstruct vacuum metric perturbations in a radiation gauge from $\psi_0$ or $\psi_4$.

Our strategy will be first to solve the Teukolsky equation numerically by separation of variables, and then obtain the metric perturbations using the CCK reconstruction procedure. This technique was pioneered by the group of Friedman \emph{et al} in Milwaukee \cite{Keidl:2006wk,Keidl:2010pm,Shah:2010bi,Shah:2012gu}, applying it successfully to obtain the metric perturbations produced by a mass moving on a circular equatorial orbit around a Kerr black hole \cite{Shah:2012gu}. Here we are extending these results to eccentric equatorial orbits in Kerr.

The extension to eccentric orbits brings new challenges. For one, the Teukolsky equation for the radial functions now has a non-zero source in a region extending from $\rmin$ to $\rmax$. This causes a problem because the CCK reconstruction procedure is ill-defined for non-vacuum regions \cite{Ori:2002uv}. Consequently, it cannot be applied mode-by-mode in the source region.

\begin{figure}[tb!]
\includegraphics[width=\columnwidth]{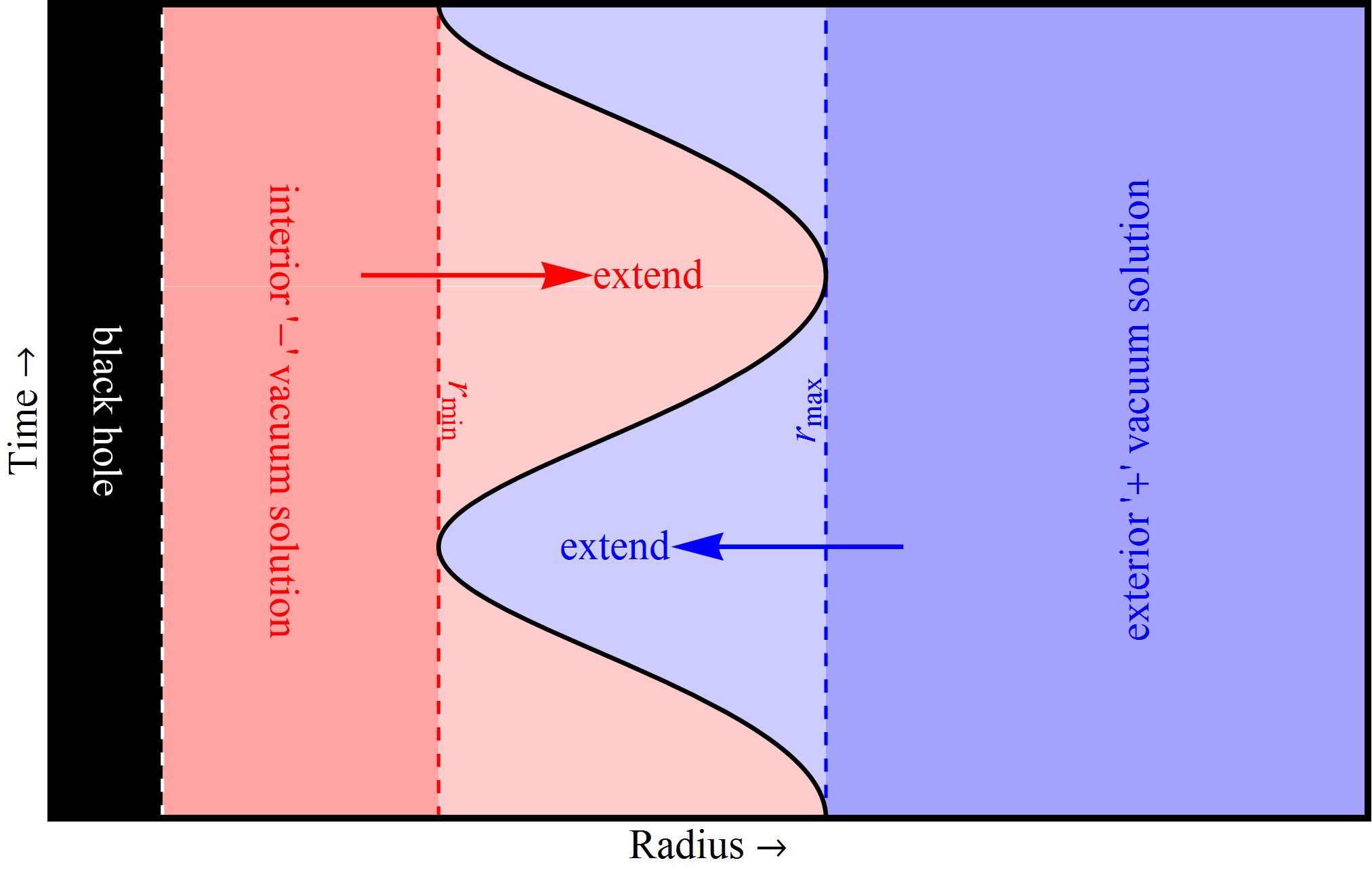}
\caption{Graphical representation of the method of extended homogeneous solutions. The field equations are solved in the interior and exterior vacuum regions in the frequency domain. To obtain the field at the particle the time domain solutions are constructed and analytically extended towards the worldline (either from the inside or outside).}\label{fig:exthom}
\end{figure}

We avoid this issue by applying the so-called ``method of extended homogeneous solutions''. This method was originally introduced \cite{Barack:2008ms} to avoid poor convergence of the mode-sum at the particle for eccentric orbits due to the Gibbs phenomenon. The basic idea is as follows. A particle traveling on an eccentric orbit naturally splits the $rt$-plane into an interior and exterior region (see Fig. \ref{fig:exthom}). Outside of the region between $\rmin$ and $\rmax$, the frequency domain equations are free of source terms and the field perturbation is given as a sum of homogeneous frequency modes with appropriate (retarded) boundary conditions at infinity (in the exterior) or at the horizon (in the interior). In the (1+1) time domain, the solution in the entire exterior/interior region is vacuum. So it can be obtained by analytically extending the frequency domain solution in the appropriate vacuum region. This analytic extension can in fact be done on a mode-by-mode basis, yielding mode-by-mode contributions to the field perturbation at the particle. The sum of these contributions (over the frequency modes) converges exponentially \cite{Barack:2008ms}.

Schematically, our approach will be as follows. For each frequency mode we solve the Teukolsky equation to find the interior and exterior vacuum solutions for $\psi_4$. Although these will only represent the physical perturbation of $\psi_4$ in the interior and exterior regions respectively, these solutions will be valid vacuum solutions everywhere. We then apply the CCK reconstruction procedure to obtain the metric perturbation corresponding to these vacuum solutions. The time domain form of the frequency domain modes of the metric perturbation are then evaluated at each point along the orbit. The full metric perturbation is  the sum of all these contributions.

The plan for the rest of this section is as follows. In section \ref{sec:teukeq}, we review the Teukolsky formalism for obtaining $\psi_4$. We pay particular attention to obtaining the asymptotic behaviour at infinity and the black hole horizon of the homogeneous solutions of the radial Teukolsky equation. The first step in the CCK procedure is to obtain the so-called Hertz potential, which itself is a solution of the Teukolsky equation with spin $s=2$. Section \ref{sec:HP} leverages the asymptotic behaviour of the solutions to the Teukolsky equation to solve the equations for the Hertz potential algebraically. The rest of the steps of the CCK procedure are discussed in section \ref{sec:MR}, where we explicitly obtain an operator that produces the (spherical harmonic) modes of $h_{uu}$ from the (spheroidal spin-weighted harmonic) modes of the Hertz potential. Section \ref{sec:comp} discusses the procedure for obtaining the remaining pieces of the metric perturbation due to perturbations to the total mass and angular momentum of the spacetime. This gives us the modes of the retarded metric perturbations, whose sum diverges at the particle. Section \ref{sec:reg} discusses the mode-sum regularization procedure for obtaining the regular ``$R$'' part of the perturbations. 

\subsection{Teukolsky Equation}\label{sec:teukeq}
We here review the Teukolksy formalism focussing on the asymptotic properties of the solutions of the Teukolsky equation. For a more comprehensive review see \cite{ST:lrr-2003-6} and the reference therein. Discussion of the static modes similar to ours can be found in \cite{Barack:1999st}.

A classical result from Teukolsky \cite{Teukolsky:1972my,Teukolsky:1973ha} is that the equations for fields $\Phi_s$ with spin-weight $s$ on a Kerr background can be solved by separation of variables. Writing\footnote{Here we anticipate that we will deal with solutions with a discrete spectrum indexed by the integers $m$ and $n$. In the more general case of fields with a continuous spectrum the sum over $n$ would be replaced by an integral over $\omega$.} 
\begin{equation}
\Phi_s =\! \frac{1}{\sqrt{2\pi}}\!\!\sum_{l,m,n}\!\! {_s\Phi_{lmn}}\R{s}{lmn}(r)\SWSH{s}{lmn}(z) \ee^{\ii(m\phi-\omega_{mn} t)},
\end{equation}
one finds that the radial mode functions $\R{s}{lmn}(r)$ and spheroidal mode functions $\SWSH{s}{lmn}(z)$ satisfy individual (uncoupled) ordinary differential equations,
\begin{align}
\begin{split}
\cbB{
	&\Delta^{-s}\d{}{r}\Bh{\Delta^{s+1}\d{}{r}}
	+\frac{K_{mn}^2-2\ii s (r-1)K_{mn}}{\Delta}
	\\
	&+4\ii s\omega_{mn} r -\TEV{s}{lmn}
}\R{s}{lmn}(r) 
= \T{s}{lmn}(r),
\end{split}\label{eq:radteukeq}\\
\begin{split}
\cbB{
	\d{}{z}\Bh{(1-z^2)\d{}{z}}
-	\frac{(m+sz)^2}{1-z^2}
+	(a\omega_{mn} z-s)^2
\\
-	s(s-1)
+	\SEV{s}{lmn}
}\SWSH{s}{lmn}(z) = 0,
\end{split}\label{eq:SWSHeq}
\end{align}
where
\begin{align}
 K_{mn} 			&\equiv (r^2+a^2)\omega_{mn}- a m,\\
 \TEV{s}{lmn} &\equiv \SEV{s}{lmn} + a^2\omega_{mn}^2-2ma\omega_{mn}.
\end{align}
The polar mode functions $\SWSH{s}{lmn}(z)$ are known as spin-weighted spheroidal harmonics. In the limit $a\omega_{mn}\to 0$, the eigenvalues reduce to 
\begin{equation}
\SEV{s}{lmn}=\TEV{s}{lmn}=l(l+1)-s(s+1),
\end{equation}
and equation \eqref{eq:SWSHeq} reduces to the familiar equation for spin-weighted spherical harmonics. In this paper we choose to normalize the spin-weighted spheroidal harmonics such that
\begin{equation}
\int\limits_{-1}^1\!\!\id{z}\SWSH{s}{lm\omega}(z)^2 =1.
\end{equation}
For our purpose, the most relevant examples of spin-weighted fields are the linear perturbations of the Weyl scalars $\Phi_2=\psi_0$ and $\Phi_{-2}=\rho^{-4}\psi_4$. The rest of this section will deal with the properties of the solutions of the radial Teukolsky equation \eqref{eq:radteukeq} with $\abs{s}=2$. For reasons that will become apparent in Sec. \ref{sec:HP}, we will focus on determining the asymptotic  behaviour of the homogeneous solutions at infinity and near the horizon.

We will distinguish three separate cases depending on whether $\omega_{mn}$ and/or $ma$ vanish.

\subsubsection{Homogeneous solutions (generic case \texorpdfstring{$\omega_{mn}\neq0$}{w\_mn!=0} and \texorpdfstring{$ma\neq0$}{ma!=0})}
In the most generic case, where neither $\omega_{mn}$ nor $ma$ vanish, the homogeneous solutions of the radial Teukolsky equation \eqref{eq:radteukeq} are oscillatory. One straightforwardly establishes that at infinity and near the horizon the homogeneous solutions behave as \cite{Teukolsky:1974yv},
\begin{alignat}{5}
\label{eq:asympinfG}
\R{s}{lmn}(r)  &= 	
\begin{aligned}[t]
&\Ainf{s}{lmn} r^{-1}\exp(-\ii\omega_{mn} r^{*})\\
 +&\Binf{s}{lmn} r^{-(2s+1)}\exp(\ii\omega_{mn} r^{*})
\end{aligned}
					&&\text{$r\to\infty$},\\
\label{eq:asymphorG}
\R{s}{lmn}(r) &=
\begin{aligned}[t]
 	&\A{s}{lmn} \Delta^{-s}\exp(-\ii k_{mn} r^{*})\\
 	+&\B{s}{lmn} \exp(\ii k_{mn} r^{*})
\end{aligned} 	
\quad&&\text{$r\to r_{+}$},
\end{alignat}
where $k_{mn} = \omega_{mn} -\frac{m a}{2r_{+}}$, and $ r^{*}= r +  \frac{2r_{+}}{r_{+}-r_{-}}\log{\frac{r-r_{-}}{2}}-\frac{2r_{-}}{r_{+}-r_{-}}\log{\frac{r-r_{+}}{2}}$ is the tortoise coordinate.

The positive and negative frequency modes ($\pm\omega_{mn}$, or $\pm k_{mn}$) are naturally identified as representing incoming and outgoing waves at infinity and the horizon. The physical solution at infinity, $\R[\pI]{s}{lmn}$, is the solution with no waves coming in from infinity, implying $\Ainf[\pI]{s}{lmn}=0$. We further normalize this solution by setting $\Binf[\pI]{s}{lmn}=1$. Similarly, the physical solution at the horizon, $\R[\pH]{s}{lmn}$, is the solution with no waves coming ``up'' the black hole horizon, leading to the conditions  $\B[\pH]{s}{lmn}=0$ and $\A[\pH]{s}{lmn}=1$. For later reference the complementary ``unphysical'' solutions at infinity and the horizon are denoted $\R[\uI]{s}{lmn}$ (defined by $\Binf[\uI]{s}{lmn}=0$ and $\Ainf[\uI]{s}{lmn}=1$) and $\R[\uH]{s}{lmn}$ (defined by $\A[\uH]{s}{lmn}=0$ and $\B[\uH]{s}{lmn}=1$).

Since \eqref{eq:radteukeq} is a second order linear differential equation, its space of homogeneous solutions is  two dimensional. Consequently, any of the four solutions $\R[\pIH]{s}{lmn}$ and $\R[\uIH]{s}{lmn}$ can be written as a linear combination of two other solutions. In particular, we can write the ``unphysical'' solutions as a linear combination of the ``physical'' solutions
\begin{align}\label{eq:genIdecomp}
\R[\uI]{s}{lmn} &=
 \frac{
\R[\pH]{s}{lmn} - \Binf[\pH]{s}{lmn}\R[\pI]{s}{lmn}
}{
\Ainf[\pH]{s}{lmn}
},
\\
\label{eq:genHdecomp}
\R[\uH]{s}{lmn} &= \frac{
\R[\pI]{s}{lmn} - \A[\pI]{s}{lmn}\R[\pH]{s}{lmn}
}{
\B[\pI]{s}{lmn}
}.
\end{align}

\subsubsection{Axisymmetric static modes (\texorpdfstring{$\omega_{mn}=0$}{w\_mn=0} and \texorpdfstring{$ma=0$}{ma=0})}
The asymptotic behaviour of the homogeneous solutions changes in the special (static) case that $\omega_{mn}$ vanishes. In that case \eqref{eq:radteukeq} becomes
\begin{equation}\label{eq:Teqx}
\begin{split}
0&=R''(x) +\frac{(1+s)(1+2x)}{x(1+x)} R'(x) +
 \\
 &\Bh{
 \frac{ \sigma^2 - 2 \ii s (1 + 2 x) \sigma }{4\bh{x(1+x)}^2}
 -\frac{(l - s) (1 + l + s) }{x(1+x)}
 } R(x) ,
\end{split}
\end{equation}
where $x=\frac{r-r_{+}}{2\kappa}$, $\sigma=-\frac{ m a}{\kappa}$, and $\kappa=	\sqrt{1-a^2}$.

This equation has explicit analytic solutions. When the product $ma$ also vanishes (i.e. for general static modes in Schwarzschild ($a=0$) or axisymmetric ($m=0$) static modes in Kerr), eq. \eqref{eq:Teqx} is solved by
\begin{equation}\label{eq:statsol1}
\begin{split}
\R{s}{lmn} 	= &\A{s}{lmn} \hh{x(1+x)}^{-\tfrac{s}{2}} \lP{l}{s}{1+2x}\\
			&+ \B{s}{lmn} \hh{x(1+x)}^{-\tfrac{s}{2}} \lQ{l}{s}{1+2x},
\end{split}
\end{equation}
where $\lP{n}{m}{x}$ and $\lQ{n}{m}{x}$ are associated Legendre P and Q functions.

These solutions have the asymptotic form,
\begin{equation} 
\begin{split}
\R{s}{lmn}(x&) =\\	
&\begin{aligned}[t]
	x^{-s} &\A{s}{lmn}\bh{\ph{1-s}{s} + \ph{2-s}{s-1}l(l+1) x \\
	 &+\ph{3-s}{s-2}\ph{l-1}{4} x^2 +\bigO(x^3)}\\
	+ &\B{s}{lmn} \hh{\text{divergent terms}}
\end{aligned}\\
&\text{as $x\to 0$,}
\end{split}
\end{equation}
and
\begin{equation}
\begin{split}
\R{s}{lmn}(x) &= \\
&\begin{aligned}[t]
 &x^{l-s}
	\Bh{\A{s}{lmn} \frac{(2l)!}{(l)!(l-s)!}+\bigO(\tfrac{1}{x})}+\\
 &x^{-l-s-1} \Bh{X\A{s}{lmn}\\
 &-\B{s}{lmn}\frac{\sqrt{\pi} s \ph{l+\frac{3}{2}}{-s-\frac{1}{2}}}{2^{2l+1}}+\bigO(\tfrac{1}{x})}
\end{aligned}
\\
&\text{as $x\to \infty$,}
\end{split}
\end{equation}
where $X$ is an unspecified function of $s$ and $l$ that we will not need. Note the use of the Pochhammer symbol
\begin{equation}
\ph{x}{n} = \frac{\Gamma(x+n)}{\Gamma(x)} = x(x+1)\cdots(x+n-1).
\end{equation}

The physical solutions are identified by imposing regularity at the horizon and infinity. At first glance this may seem impossible, since both solutions in \eqref{eq:statsol1} appear divergent at the horizon. This is due to irregularity of the Kinnersley tetrad at the horizon, and the physical regularity condition is that $\Delta^s R(r)$ should be smooth at the horizon \cite{Barack:1999st}. We thus identify the physical solution at infinity, $\R[\pI]{s}{lmn}$, as set by the conditions 
\begin{align}
\A[\pI]{s}{lmn} &=0\quad\text{and}\quad\\
\B[\pI]{s}{lmn} &=(-1)^{s+1}2 s (2\kappa)^{-s} \frac{(l-s)!}{(l+s)!},
\end{align}and the physical solution at the horizon, $\R[\pH]{s}{lmn}$, as set by 
\begin{equation}
\A[\pH]{s}{lmn} =(2\kappa)^{-s}\quad\text{and}\quad\B[\pH]{s}{lmn} =0.
\end{equation}

\subsubsection{Static modes (\texorpdfstring{$\omega_{mn}=0$}{w\_mn=0} and \texorpdfstring{$ma\neq0$}{ma!=0})}
We finally consider the very special case that $\omega_{mn}=0$ while both $a$ and $m$ are non-zero. For modes sourced by a particle moving on an equatorial orbit this can only happen if an integer combination of the radial and azimuthal frequencies $\Omega_r$ and $\Omega_\phi$ vanish. Such special orbits are known as $r\phi$-resonances, which can cause a coherent build up of linear momentum flux to infinity \cite{Meent:2014}, but should not have a direct impact on the local dynamics of the binary system at leading order in $\mr$.

The most general solution in the case that $ma\neq 0$  is given by,
\begin{equation}
\begin{split}
\R{s}{lmn} &=  \A{s}{lm} \hh{\frac{1+x}{x}}^{\ii\frac{\sigma}{2}} \hh{x(1+x)}^{-s}\\
 &\times\F(-l-s;1+l-s;1-s-\ii\sigma;-x)\\
 &+\B{s}{lm} \hh{\frac{1+x}{x}}^{-\ii\frac{\sigma}{2}}\\
 &\times\F(-l+s;1+l+s;1+s+\ii\sigma;-x),
\end{split}
\end{equation}
where the $\F$ are hypergeometric functions.

Near infinity and the horizon this solution has the following asymptotic form
\begin{align}
\begin{split}
\R{s}{lmn}(x) &=  
x^{-s-\ii\frac{\sigma}{2}} \hh{\A{s}{lm} +\bigO(x)}\\ 
&\quad+  x^{\ii\frac{\sigma}{2}} \hh{\B{s}{lm} +\bigO(x)}\\
&\text{as $x\to 0$},
\end{split}
\\
\intertext{and}
\begin{split}
\R{s}{lmn}(x) &= x^{l-s} \Bh{
	\A{s}{lm} \frac{\ph{1+l+s}{l-s}}{\ph{1-s-\ii\sigma}{l+s}}
	\\
	 &+\B{s}{lm}\frac{\ph{1+l-s}{l+s}}{\ph{1+s+\ii\sigma}{l-s}}+\bigO(\tfrac{1}{x})}
	 \\
  &+x^{-l-s-1} \hh{X\A{s}{lm}+Y\B{s}{lm}+\bigO(\tfrac{1}{x})}
  \\
&\text{as $x\to \infty$},
\end{split}
\end{align}
where $X$ and $Y$ are unspecified functions of $s$, $l$, and $\sigma$ that we will not need.

Again imposing regularity, we identify the physical solutions at infinity $\R[\pI]{s}{lmn}$ by 
\begin{align}
\A[\pI]{s}{lmn} &=-(2\kappa)^{-s}\frac{(l-s)!\ph{1-s-\ii\sigma}{l+s}}{(l+s)!\ph{1+s+\ii\sigma}{l-s}},\text{and}\\
\B[\pI]{s}{lmn} &=1,
\end{align} and the physical solution at the horizon, $\R[\pH]{s}{lmn}$, by 
\begin{equation}
\A[\pH]{s}{lmn} =(2\kappa)^{-s} \quad\text{and}\quad \B[\pH]{s}{lmn} =0.
\end{equation}
Moreover, we will need the irregular solution at the horizon $\R[\uH]{s}{lmn}$ defined by
\begin{equation}
\A[\uH]{s}{lmn} =0 \quad\text{and}\quad \B[\uH]{s}{lmn} =(2\kappa)^{-s},
\end{equation}
which can be written as a linear combination of the physical solutions at the horizon and infinity,
\begin{align}\label{eq:statudecomp}
\R[\uH]{s}{lmn} &=  \frac{
\R[\pI]{s}{lmn} - \A[\pI]{s}{lmn}\R[\pH]{s}{lmn}
}{
\B[\pI]{s}{lmn}
}\\
&= \R[\pI]{s}{lmn} + \frac{(l-s)!\ph{1-s-\ii\sigma}{l+s}}{(l+s)!\ph{1+s+\ii\sigma}{l-s}} \R[\pH]{s}{lmn}
.
\end{align}

\subsubsection{Inhomogeneous modes}
The solution of the inhomogeneous Teukolsky equation, sourced by a particle moving on an equatorial orbit, in the vacuum regions  inside ($\pH$) and outside ($\pI$) the orbit can be written as a linear combination of the homogeneous solutions,
\begin{equation}\label{eq:psi4exp}
\psi_{4}^\pm = \frac{\rho^{4}}{\sqrt{2\pi}}\sum_{l,m,n} Z_{lmn}^\pm \R[\pm]{-2}{lmn}(r)\; \SWSH{-2}{lmn}(z)\ee^{\ii(m\phi-\omega_{mn} t)}.
\end{equation}
The coefficients $Z_{lm\omega}^\pm$ can be determined using variations of parameters,
\begin{equation}\label{eq:sourceint}
Z_{lmn}^\pm = \int_{r_{\mathrm{min}}}^{r_{\mathrm{max}}}  \frac{
	\R[\mp]{-2}{lmn}(r)\T{-2}{lmn}(r)
	}{
	W[\R[\pI]{-2}{lmn},\R[\pH]{-2}{lmn}](r)
	}
	\id{r},
\end{equation}
where $W[f_1,f_2]$ denotes the Wronskian of the solutions $f_1$ and $f_2$. An explicit expression for the ``source'' term $\T{-2}{lmn}(r)$ is given in \cite{Meent:2015a}.

\subsection{Hertz Potential}\label{sec:HP}
The reconstruction of the metric perturbations in the CCK formalism is defined in terms of the so-called Hertz potential. In vacuum regions, the Hertz potential for metric perturbations in the outgoing radiation gauge $\Psi_{ORG}$ satisfies two conditions \cite{Lousto:2002em}. First,  $\Psi_{ORG}$ satisfies the Teukolksy equation for $s=2$ fields in vacuum. Second, $\Psi_{ORG}$ satisfies a fourth order differential equation with $\psi_4$ acting as a source term,
\begin{equation}\label{eq:psiORG}
\frac{1}{32}\Delta^2 (D_0^{\dagger})^4 \Delta^2 \bar{\Psi}_\textrm{ORG} = \rho^{-4}\psi_4,
\end{equation}
where $D_0^{\dagger}= \partial_r- \frac{(r^2+a^2)\partial_t+a\partial_\phi}{\Delta}$, and the over bar denotes complex conjugation.
 
The first condition implies that $\Psi_{ORG}$ in the interior and exterior vacuum regions can be decomposed in spin-weighted spheroidal harmonics,
\begin{equation}\label{eq:PsiExp}
\Psi_{ORG}^\pm = \frac{1}{\sqrt{2\pi}}\sum_{l,m,n} \Psi_{lmn}^\pm \R[\pm]{2}{lmn}(r)\; \SWSH{2}{lmn}(z)\ee^{\ii(m\phi-\omega_{mn} t)}.
\end{equation}
Furthermore, the linear operator on the left hand side in \eqref{eq:psiORG} neatly separates over the spheroidal modes, with the operator acting on the radial modes given by
\begin{equation}
D_{mn}^\dagger =\partial_r +\ii\frac{\omega_{mn}(r^2+a^2)-ma}{\Delta} \equiv D_{-m-n}.
\end{equation}
Consequently, \eqref{eq:psiORG} can be solved mode-by-mode. In particular, we observe that the resulting operator acting on the radial modes is a linear operator that maps solutions of the radial (vacuum) Teukolsky equation with $s=2$ into solutions of the radial (vacuum) Teukolsky equation with $s=-2$. Consequently, if we fix a basis on the (2-dimensional) space of solutions of the (vacuum) Teukolsky equation (as we did in section \ref{sec:teukeq}), the action of this operator is given by a 2-by-2 matrix, which can easily be inverted. The basis in section \ref{sec:teukeq} was chosen to ensure that this matrix is diagonal, simplifying the inversion. Moreover, the action of the radial operator can be determined at any radius $r$. In particular, we can determine the action (and its inverse) in the asymptotic regions at infinity and near the horizon, requiring only the asymptotic behaviour of the homogeneous solutions as obtained in the previous section.

Ori \cite{Ori:2002uv} performed this analysis for the relation between (generic) modes of the incoming radiation gauge (IRG) Hertz potential and $\psi_0$. We here extend that analysis to the ORG Hertz potenital, $\Psi_{ORG}$, and $\psi_4$, also including all the static modes.

Expanding the left hand side of \eqref{eq:psiORG} using \eqref{eq:PsiExp} we find
\begin{multline}
\frac{1}{32}\Delta^2 (D_0^\dagger)^4 \Delta^2 \bar\Psi^\pm_{ORG}
 = \sum_{l,m,n} \frac{\bar{\Psi}_{lmn}^\pm}{32}  
 						\Delta^2 (D_{-m-n}^\dagger)^4\Delta^2
\\ 						
 						\times\Rb[\pm]{2}{lmn}(r)\;
 						\SWSH{2}{lmn}(z)\ee^{-\ii(m\phi-\omega_{mn} t)}.
\end{multline}
Relabelling $(m,n)$ to $(-m,-n)$ and using the identities $\SWSH{s}{lmn}(z)=(-1)^{s+m}\SWSH{-s}{l-m-n}(z)$ and $\Rb{s}{lmn}=\R{s}{l-m-n}$, this becomes
\begin{multline}
\frac{1}{32}\Delta^2 (D_0^\dagger)^4 \Delta^2 \bar\Psi^\pm_{ORG} = \sum_{l,m,n} (-1)^{m} \frac{\bar{\Psi}_{l-m-\omega}^\pm}{32}\times\\
						\Delta^2 (D_{mn}^\dagger)^4\Delta^2\R[\pm]{2}{lmn}(r)\;
						\SWSH{-2}{lmn}(z)\ee^{\ii(m\phi-\omega_{mn} t)}.
\end{multline}
Comparing with the expansion \eqref{eq:psi4exp} of the right hand side of \eqref{eq:psiORG}, and noting the orthogonality of the spheroidal and Fourier modes, we obtain,
\begin{equation}\label{eq:modeinv1a}
\begin{split}
Z_{lmn}^\pm \R[\pm]{-2}{lmn}(r)&=\\
\frac{(-1)^{m}}{32} &\bar{\Psi}_{l-m-n}^\pm \Delta^2 (D_{mn}^\dagger)^4\Delta^2\R[\pm]{2}{lmn}(r)
.
\end{split}
\end{equation}
By the Teukolsky-Starobinsky identities (see e.g. \cite{ChandraBook} section 81) this is equivalent to
\begin{equation}\label{eq:modeinv2a}
\begin{split}
 Z_{lmn}^\pm(D_{mn})^4 \R[\pm]{-2}{lmn}(r)&=\\
\frac{(-1)^{m}}{32} &\bar{\Psi}_{l-m-n}^\pm p_{lmn}\R[\pm]{2}{lmn}(r),
\end{split}
\end{equation}
where
\begin{equation}
\begin{split}
p_{lmn} = &\bh{(\TEV{-2}{lmn} + 2)^2 + 4 m a \omega_{mn} - 
     4 a^2 \omega_{mn}^2}
     \\
     &\times\bh{\TEV{-2}{lmn}^2 + 36 m a \omega_{mn} - 
     36 a^2  \omega_{mn}^2}
     \\
     &+ (2 \TEV{-2}{lmn} + 3) (96 a^2 \omega_{mn}^2 - 
     48 m a \omega_{mn})\\
     &+ 144 \omega_{mn}^2 (1 - a^2).
\end{split}
\end{equation}
We now observe that if $f(r)$ solves the radial Teukolsky equation \eqref{eq:radteukeq} with spin $s$, then 
$\Delta^s\bar{f}(r)$ solves the radial Teukolsky equation with the opposite spin $-s$, as can easily be verified by inserting $\Delta^s\bar{f}(r)$ in \eqref{eq:radteukeq} with spin $-s$. For generic ($\omega\neq 0$) modes, the asymptotic behaviour tells us that this relation exchanges physical (retarded) for unphysical (advanced) boundary conditions,
\begin{equation}\label{eq:gentrans}
\Delta^s \Rb[\pIH]{s}{lmn} = \R[\uHI]{-s}{lmn}.
\end{equation}
Using this relation Eqs. \eqref{eq:modeinv1a} and \eqref{eq:modeinv2a} become
\begin{equation}\label{eq:modeinv1}
\begin{split}
 Z_{lmn}^\pm \R[\pm]{-2}{lmn}(r)&=\\
 \frac{(-1)^{m}}{32} &\bar{\Psi}_{l-m-n}^\pm \Delta^2 (D_{mn}^\dagger)^4\R[\not{\pm}]{-2}{l-m-n}(r)
 \end{split}
\end{equation}
and
\begin{equation}\label{eq:modeinv2}
\begin{split}
Z_{lmn}^\pm (D_{mn})^4 \Delta^2 \R[\pm]{-2}{lmn}(r)&=\\
\frac{(-1)^{m}}{32} \bar{\Psi}_{l-m-n}^\pm &p_{lmn}\R[\not\pm]{-2}{l-m-n}(r)
 .
 \end{split}
\end{equation}
If we insert the asymptotic expansion \eqref{eq:asympinfG} into \eqref{eq:modeinv1} we find
\begin{equation}
\begin{split}
 Z_{lmn}^\pI r^{3}\exp(\ii\omega_{mn} r^{*})&=\\
 \frac{(-1)^{m}}{32}&\bar{\Psi}_{l-m-n}^\pI 16\omega^4 r^{3}\exp(\ii\omega_{mn} r^{*})
 ,
 \end{split}
\end{equation}
for the solution at infinity. This can easily be solved for $\Psi^\pI_{lmn}$,
\begin{equation}
\Psi^\pI_{lmn} = (-1)^{l+m} 2\frac{Z_{lmn}^\pI}{\omega_{mn}^{4}},
\end{equation}
where we used that $\bar{Z}_{lmn} = (-1)^l Z_{l-m-n}$. 

However, for the physical mode at the horizon if one inserts the asymptotic expansion near the horizon \eqref{eq:asymphorG} into \eqref{eq:modeinv1}, one finds that the leading terms cancels, yielding no useful information. Using \eqref{eq:modeinv2} instead one finds,
\begin{align}
\begin{split}
 Z_{lmn}^\pH (2\kappa)^4\ph{\ii(\sigma+2\omega_{mn})-2}{4} \exp(-\ii k_{mn} r^{*})&=\\
 \frac{(-1)^{m}}{32}\bar{\Psi}_{l-m-n}^\pH p_{lmn} \exp(-\ii k_{mn} r^{*}),
\end{split}
\end{align}
which we can solve for $\Psi^\pH_{lmn}$,
\begin{equation}
\Psi^\pH_{lmn} = (-1)^{l+m} \frac{32 Z_{lmn}^\pH }{ p_{lm\omega} }  (2\kappa)^4\ph{\ii(\sigma+2\omega_{mn})-2}{4}.
\end{equation}

For the static case $\omega=0$, but $ma\neq 0$, one easily verifies that
\begin{align}\label{eq:stat1Itrans}
\Delta^s \Rb[\pI]{s}{lmn} &=  \frac{(l-s)!\ph{1-s-\ii\sigma}{l+s}}{(l+s)!\ph{1+s+\ii\sigma}{l-s}} \R[\pI]{-s}{lmn},
\intertext{and}
\label{eq:stat1Htrans}
\Delta^s \Rb[\pH]{s}{lmn} &= \R[\uH]{-s}{lmn}.
\end{align}
Repeating the analysis above we find
\begin{align}
\Psi_{lmn}^\pI	
 &=
 (-1)^{l+m}  \frac{32}{\ph{\ii\sigma-2}{4}}Z_{lmn}^\pI,
\\
\Psi_{lmn}^\pH	
 &=
 (-1)^{l+m} 32 \frac{\ph{\ii\sigma-2}{4}}{\bh{\ph{l-1}{4}}^2} Z_{lmn}^\pH.
\end{align}
Similarly, for $\omega=ma=0$,
\begin{align}
\label{eq:stat2Itrans}
\Delta^s \Rb[\pI]{s}{lmn} &= - \frac{(l-s)!}{(l+s)!} \R[\pI]{-s}{lmn}\text
{, and}\\
\label{eq:stat2Htrans}
\Delta^s \Rb[\pH]{s}{lmn} &=  \frac{(l+s)!}{(l-s)!} \R[\pH]{-s}{lmn},
\end{align}
leading to
\begin{align}
\Psi_{lmn}^\pI	
 &=
 (-1)^{l+m+1} 32 Z_{lmn}^\pI,
\\
\Psi_{lmn}^\pH	
 &=
(-1)^{l+m}  \frac{32}{\hh{\ph{l-1}{4}}^2} Z_{lmn}^\pH.
\end{align}

This brings us to the main result from this section: a set of algebraic expressions for the asymptotic amplitudes $\Psi_{lmn}^\pIH$ of the outgoing radiation gauge Hertz potential in terms of the asymptotic amplitudes $Z_{lmn}^\pIH$ of $\psi_4$. In summary, these expression are,
\begin{widetext}\begin{align}\label{eq:psiI}
\Psi_{lmn}^\pI &=\begin{cases}
(-1)^{l+m}\frac{2}{\omega_{mn}^{4}}Z_{lmn}^\pI
\hspace{104pt}\phantom{.}&\text{for $\omega_{mn}\neq 0$,}
\\
(-1)^{l+m} \frac{32}{\ph{\ii\sigma-2}{4}}Z_{lmn}^\pI
&\text{for $\omega_{mn}= 0$ but $ma\neq 0$,}
\\
(-1)^{l+m+1} 32  Z_{lmn}^\pI
&\text{for $\omega_{mn}=ma= 0$,}
\end{cases}
\\
\label{eq:psiH}
\Psi_{lmn}^\pH &=\begin{cases}
(-1)^{l+m} \frac{32}{ p_{lm\omega} } (2\kappa)^4\ph{\ii(\sigma+2\omega_{mn})-2}{4} Z_{lmn}^\pH 
&\text{for $\omega_{mn}\neq 0$,}
\\
(-1)^{l+m} 32\frac{\ph{\ii\sigma-2}{4}}{\bh{\ph{l-1}{4}}^2} Z_{lmn}^\pH
&\text{for $\omega_{mn}= 0$ but $ma\neq 0$,}
\\
(-1)^{l+m}  \frac{32}{\hh{\ph{l-1}{4}}^2} Z_{lmn}^\pH
&\text{for $\omega_{mn}=ma= 0$.}
\end{cases}
\end{align}
In addition, by combining \eqref{eq:genIdecomp}, \eqref{eq:genHdecomp}, \eqref{eq:statudecomp}, \eqref{eq:gentrans}, \eqref{eq:stat1Itrans}, \eqref{eq:stat1Htrans}, \eqref{eq:stat2Itrans}, and \eqref{eq:stat2Htrans}, we can obtain the $s=2$ homogeneous solutions of the radial Teukolsky equation needed in \eqref{eq:PsiExp} explicitly in terms of the $s=-2$  homogeneous solutions,
\begin{align}\label{eq:R2I}
\R[\pI]{2}{lmn}(r) &=\begin{cases}
\frac{1}{\Delta^2}
\hh{
\frac{1}{\Ainfb[\pH]{-2}{lmn}}\Rb[\pH]{-2}{lmn}(r)
 - \frac{\Binfb[\pH]{-2}{lmn}}{\Ainfb[\pH]{-2}{lmn}}\Rb[\pI]{-2}{lmn}(r)
}
&\text{for $\omega_{mn}\neq 0$,}
\\
\frac{
\ph{1-2+\ii\sigma}{l+2}
}{
\Delta^{2}\ph{l-1}{4}\ph{1+2-\ii\sigma}{l-2}
} \Rb[\pI]{-2}{lmn}(r)
&\text{for $\omega_{mn}= 0$ but $ma\neq 0$.}
\\
\frac{
-1 
}{
\Delta^{2}\ph{l-1}{4}
}
\Rb[\pI]{-2}{lmn}
&\text{for $\omega_{mn}=ma= 0$,}
\end{cases}
\\
\label{eq:R2H}
\R[\pH]{2}{lmn}(r) &=\begin{cases}
\frac{1}{\Delta^2}
\hh{
\frac{1}{\Bb[\pI]{-2}{lmn}}\Rb[\pI]{-2}{lmn}(r) 
- \frac{\Ab[\pI]{-2}{lmn}}{\Bb[\pI]{-2}{lmn}}\Rb[\pH]{-2}{lmn}(r)
}
&\text{for $\omega_{mn}\neq 0$,}
\\
\frac{
1
}{
\Delta^{2}
}\hh{\Rb[\pI]{-2}{lmn} + \frac{
\ph{l-1}{4}\ph{1+2+\ii\sigma}{l-2}
}{
\ph{1-2-\ii\sigma}{l+2}
} \Rb[\pH]{-2}{lmn}}
&\text{for $\omega_{mn}= 0$ but $ma\neq 0$,}
\\
\frac{
\ph{l-1}{4}
}{
\Delta^{2}
} \Rb[\pH]{-2}{lmn}
&\text{for $\omega_{mn}=ma= 0$.}
\end{cases}
\end{align}
\end{widetext}
Consequently, once we have obtained the mode expansion \eqref{eq:psi4exp} for $\psi_4$, we can use these algebraic relations to obtain $\Psi_{ORG}$ (in the interior and exterior vacuum  regions) without solving any further differential equations.

\subsection{Metric Reconstruction}\label{sec:MR}
The next step in the CCK procedure is to produce the reconstructed metric components from the Hertz potential. These will be produced in an outgoing radiation gauge defined by
\begin{alignat}{3}
h_{2a}&\equiv \tet{2}{\mu}\tet{a}{\nu}h_{\mu\nu} &&=0,\\
h_{34}&\equiv \tet{3}{\mu}\tet{4}{\nu}h_{\mu\nu} &&=0.
\end{alignat}
The CCK procedure obtains the remaining non-zero components of the metric perturbation by acting on $\Psi_{ORG}$ with certain second order linear differential operators (see e.g. \cite{Lousto:2002em}). With our sign conventions these are given by
\begin{alignat}{3}
h_{11} &\equiv \tet{1}{\mu}\tet{1}{\nu}h_{\mu\nu} &&= \MROP{11}\Psi_{ORG}+ c.c.,
\\
h_{13} &\equiv \tet{1}{\mu}\tet{3}{\nu}h_{\mu\nu} &&= \MROP{13}\Psi_{ORG},\text{ and}
\\
h_{33} &\equiv \tet{3}{\mu}\tet{3}{\nu}h_{\mu\nu} &&= \MROP{33}\Psi_{ORG},
\end{alignat}
with
\begin{align}
\MROP{11} &= \rho^{-4}\hh{\bar\delta-3\alpha-\bar\beta+5\pi}\hh{\bar\delta-4\alpha+\pi},
\displaybreak[0]\\
\MROP{13} &= \frac{\rho^{-4}}{2}\cbB{
	\bh{\bar\delta-3\alpha+\bar\beta+5\pi+\bar\tau}\bh{\hat\Delta+\mu-4\gamma}
	\\
	&\quad+
	\bh{\hat\Delta+5\mu-\bar\mu-3\gamma-\bar\gamma}\bh{\bar\delta-4\alpha+\pi}
	 },\text{ and}\nonumber
\\
\MROP{33} &= \rho^{-4}\hh{\hat\Delta+5\mu-3\gamma+\bar\gamma}\hh{\hat\Delta+\mu-4\gamma}.
\end{align}
Here $\bar\delta = \tet{4}{\mu}\partial_\mu$, $\hat\Delta = \tet{2}{\mu}\partial_\mu$, and the remaining Greek letters are the usual spin-coefficients in the Newman-Penrose formalism. (Explicit expressions consistent with our conventions are given in \cite{Meent:2015a}). 

Our goal in the rest of this section is to construct $h_{uu}$ from $\Psi_{ORG}$,
\begin{equation}\label{eq:huulim}
h_{uu}^{\pIH}=\lim_{x \to x_0^\pIH}u^a u^b h_{ab} = \lim_{x \to x_0^\pIH} \MROP{uu}\Psi_{ORG}^\pIH +c.c.
\end{equation}
with
\begin{equation}
\MROP{uu} =u^1u^1\MROP{11}+2 u^1 u^3 \MROP{13} +u^3u^3\MROP{33},
\end{equation}
and where the limit $x\to x_0^\pIH$ symbolizes approaching the particle worldline from either the exterior ($\pI$) or interior ($\pH$) vacuum region. As written here Eq. \eqref{eq:huulim} makes little sense, since the particle four-velocity $u^\mu$ is not a field, but a quantity defined only on the particle worldline. To make sense of the limiting procedure we must extend $u^\mu$ to a suitably smooth field defined in a neighbourhood of the worldline. Following the literature \cite{Heffernan:2012vj,Barack:2009ux}, we choose a ``rigid'' extension, i.e. we choose to extend the four-velocity $u^\mu$ such that 
\begin{equation}
u^\mu(t_0,r,z,\phi)=u^\mu(t_0,r_0,z_0,\phi_0),
\end{equation}
using that the particle crosses each $t_0$-slice at a unique point $(r_0,z_0,\phi_0)$.

A second issue with \eqref{eq:huulim} is that the limit towards the particle worldline will obviously diverge. However, to obtain $\Delta U$ we only need the regular ``R'' part of $h_{ab}$. To obtain the regular part, we will utilize the mode-sum formalism \cite{Barack:2001gx,Barack:2002bt,Barack:2002mh}. This requires that we expand $h_{uu}$ in spherical harmonics, whereas we obtain $\Psi_{ORG}$ expanded in spin-weighted spheroidal harmonics. The remainder of this section will be devoted to explicitly evaluating the action of $\MROP{uu}$ and ``re-expanding''  the result in spherical harmonics to obtain the ``$l$-modes''.

We start by inserting the expansion of the Hertz potential \eqref{eq:PsiExp} in \eqref{eq:huulim},
\begin{equation}\label{eq:huu1}
\begin{split}
h_{uu}^\pIH = \lim_{x \to x_0^\pIH} \frac{1}{\sqrt{2\pi}}&\sum_{l,m,n} \MROP{uu}\cbB{\Psi_{lmn}^\pIH \R[\pIH]{2}{lmn}(r) 
\\
&\times\SWSH{2}{lmn}(z)\ee^{\ii(m\phi-\omega_{mn} t)}} +c.c.
\end{split}
\end{equation}
The first step towards re-expanding the result in spherical harmonics is to expand the spin-weighted spheroidal harmonics $\SWSH{s}{lmn}(z)$ in spin-weighted spherical harmonics $\Y{s}{lm}(z)$,
\begin{equation}
\SWSH{s}{l_1mn}(z)=\sum_{l_2}\Sb{s}{mn}{l_1}{l_2} \Y{s}{l_2m}(z).
\end{equation}
Hughes \cite{Hughes:1999bq} has described a robust way of numerically obtaining the transformation matrix $\Sb{s}{mn}{l_1}{l_2}$.

After expanding to spherical harmonics, we can evaluate the action of the linear operator $\MROP{uu}$. Observing that derivatives with respect to $z$ can be rewritten in terms of the spin-weight lowering operator $\SWL{s}=\sqrt{1-z^2}\nh{\partial_z+\frac{\ii}{1-z^2}\partial_\phi+\frac{s z}{1-z^2}}$, the result has the  schematic form
\begin{equation}
\begin{split}
h_{uu}^\pIH = \lim_{x \to x_0^\pIH}\sum_{m,n} {\ee}^{\ii(m\phi-\omega_{mn} t)}\sum_{l_1,l_2,i,s}\MC_{l_1mnsi}(r,z)&
\\
\times\Psi_{l_1mn}^\pIH\R[\pIH,(i)]{2}{l_1mn}(r)\Sb{2}{mn}{l_1}{l_2} \Y{s}{l_2m}(z)+c.c&.,
\end{split}
\end{equation}
where $\MC$ is a numerical coefficient depending on the field point $(r,z)$, the particle position $(r_0,z_0)$, and the indices $(l_1,m,n,s,i)$.

Next, we expand the spin-weighted spherical harmonics $\Y{s}{lm}(z)$ to ordinary spherical harmonics by  using the following identities
\begin{align}
\Y{2}{l_1m}(z)&=\sum_{l_2}
\frac{
\YA{2}{m}{l_1}{l_2} \Y{}{l_2m}(z)
}{
1-z^2
}
,\\
\Y{1}{l_1m}(z)&=\sum_{l_2}
\frac{
\YA{1}{m}{l_1}{l_2}\Y{}{l_2m}(z)
}{
\sqrt{(l_1-1)(l_1+2)}\sqrt{1-z^2}
}
,\\
\Y{0}{l_1m}(z)&=\sum_{l_2}\frac{
\YA{0}{m}{l_1}{l_2}\Y{}{l_2m}(z)
}{
\sqrt{(l_1-1)l_1(l_1+1)(l_1+2)}
} ,
\end{align}
where (using Wigner $3j$-notation)
\begin{align}
&\begin{aligned}
\YA{2}{m}{l_1}{l_2} = 
(-1)^m\sqrt{\frac{8}{3}(2l_1+1)(2l_2+1)}&
\\
\times\begin{pmatrix}
2 & l_1 & l_2\\
0 & m	& -m
\end{pmatrix}
&\begin{pmatrix}
2 & l_1 & l_2\\
2 & -2	& 0
\end{pmatrix}
,
\end{aligned}
\\
&\begin{aligned}
\YA{1}{m}{l_1}{l_2} = 
&(-1)^{m+1}
\sqrt{2(l_1-1)(l_1+2)(2l_1+1)}
\\
&\quad\times\sqrt{2l_2+1}
\begin{pmatrix}
1 & l_1 & l_2\\
0 & m	& -m
\end{pmatrix}
\begin{pmatrix}
1 & l_1 & l_2\\
1 & -1	& 0
\end{pmatrix}
,
\end{aligned}\\
\intertext{and}
&\YA{0}{m}{l_1}{l_2} = \sqrt{(l_1-1)l_1(l_1+1)(l_1+2)}\delta_{l_1l_2}.
\end{align}
The normalization of the coefficients $\YA{s}{m}{l_1}{l_2}$ has been chosen as to cancel the $l_1$ dependence of the coefficients $\MC_{l_1mnsi}$. The result is
\begin{equation}\label{eq:huu3}
\begin{split}
h_{uu}^\pIH = \lim_{x \to x_0^\pIH}\sum_{m,n} \ee^{\ii(m\phi-\omega_{mn} t)}\sum_{i,s}\MC_{mnsi}(r,&z)
\\
\times\sum_{l_1,l_2,l_3}
\Psi_{l_1mn}^\pIH\R[\pIH,(i)]{2}{l_1mn}(r)&\Sb{2}{mn}{l_1}{l_2}
\\
\times\YA{s}{m}{l_2}{l_3}\Y{}{l_3m}(&z)+c.c.,
\end{split}
\end{equation}
where $\MC$ is now a different coefficient that (among other things) still depends on the field point $z$. Observing the identities
\begin{align}
\bar\MC_{mnsi}(r,z) &= (-1)^s\MC_{-m-nsi}(r,-z),\\
\bar\Psi_{l_1mn}^\pIH&= (-1)^l\Psi_{l_1-m-n}^\pIH,\\
\Rb[\pIH,(i)]{2}{l_1mn}(r) &= \R[\pIH,(i)]{2}{l_1-m-n}(r),\displaybreak[0]\\
\Sb{2}{mn}{l_1}{l_2}&=(-1)^{l_1+l_2}\Sb{2}{-m-n}{l_1}{l_2},\\
\YA{s}{m}{l_2}{l_3} &= (-1)^{s+l_2+l_3}\YA{s}{-m}{l_2}{l_3},\\
\Y{}{l_3m}(z) &= (-1)^m\Y{}{l_3-m}(z),
\end{align}
and relabelling appropriately, we can evaluate the complex conjugate terms,
\begin{widetext}
\begin{align}
h_{uu}^\pIH &=\lim_{x \to x_0^\pIH}\sum_{m,n} \ee^{\ii(m\phi-\omega_{mn} t)}
\!\!\sum_{\substack{i,s\\l_1,l_2,l_3}}\!\!
\bh{\MC_{mnsi}(r,z)+(-1)^{l_3+m}\MC_{mnsi}(r,-z)}\Psi_{l_1mn}^\pIH\R[\pIH(i)]{2}{l_1mn}(r)\Sb{2}{mn}{l_1}{l_2}\YA{s}{m}{l_2}{l_3}\Y{}{l_3m}(z)
\\
&=\lim_{x \to x_0^\pIH}\sum_{m,n} \ee^{\ii(m\phi-\omega_{mn} t)}
\!\!\sum_{\substack{i,s\\l_1,l_2,l_3}}\!\!
\MC_{l_3mnsi}(r,z^2)\Psi_{l_1mn}^\pIH\R[\pIH,(i)]{2}{l_1mn}(r)\Sb{2}{mn}{l_1}{l_2}\YA{s}{m}{l_2}{l_3}z^{\frac{1-(-1)^{l_3+m}}{2}}\Y{}{l_3m}(z)\label{eq:huu3b},
\end{align}
\end{widetext}
where in the last line we have redefined the coefficients $\MC$, which are now only a function of $z^2$.


To eliminate the dependence on the field point $z$ in the coefficients we utilize the freedom granted by the presence of the limit towards the worldline. Since we are only interested in the limiting value at the worldline, we are in principle free to multiply each term in the sum in \eqref{eq:huu3b} by a function $f(r,z)$ that smoothly approaches 1 at the worldline.\footnote{Although any choice is allowed a priori, the price we pay is that we have to account for the choices made in the calculation of the singular part in Sec. \ref{sec:reg}. In practice, we want to make our choice such that the leading order divergent structure of the singular field encoded in the regularization parameters is unaffected.} For each term in the expansion \eqref{eq:huu3b} we choose $f$ such that
\begin{equation}\label{eq:bext} 
f(r,z) = \frac{
\MC_{mnsi}(r_0,z_0^2)
}{
\MC_{mnsi}(r,z^2)
}.
\end{equation}
That is, we keep only the leading constant term of each of the coefficients $\MC$.

We finally rid ourselves of the final dependence on $z$ by utilizing the identity
\begin{equation}
z^j\Y{}{l_1m}(z)=\sum_{l_2}\YB{j}{m}{l_1}{l_2}
 \Y{}{l_2mn}(z),
\end{equation}
with
\begin{equation}
\begin{split}
\YB{1}{m}{l_1}{l_2} = (-1)^{m+l_1+1}&(l_1-l_2)
\\
\times&\sqrt{\frac{l_1+l_2+1}{2}}
\begin{pmatrix}
1 & l_1 & l_2\\
0 & m	& -m
\end{pmatrix},
\end{split}
\end{equation}\vspace{1em}
and
\begin{align}
\YB{j}{m}{l_1}{l_2} &= \bh{\YB{1}{m}{l_1}{l_2}}^j 
\\
&= 
\sum_{\ell_1,\ldots,\ell_{j-1}}
\YB{1}{m}{l_1}{\ell_1}
\Bh{\prod_{k=1}^{j-2}\YB{1}{m}{\ell_k}{\ell_{k+1}}}
\YB{1}{m}{\ell_{j-1}}{l_2}.
\end{align}
This allows us to completely evaluate the limit and obtain the $l$-modes of $h_{uu}$,
\begin{widetext}
\begin{align}
\label{eq:huu4}
h_{uu}^\pIH &=\frac{1}{\sqrt{2\pi}}\sum_{m,n} \ee^{\ii(m\phi_0-\omega_{mn} t_0)}
\sum_{i,s}
\sum_{
\substack{l_1,l_2,\\l_3,l_4}}\MC_{l_3mnsi}(r_0,z_0^2)\Psi_{l_1mn}^\pIH\R[\pIH,(i)]{2}{l_1mn}(r_0)\Sb{2}{mn}{l_1}{l_2}\YA{s}{m}{l_2}{l_3}\YB{\hh{\frac{1-(-1)^{l_3+m}}{2}}}{m}{l_3}{l_4}\Y{}{l_4m}(z_0)
\\
\label{eq:huu5}
&=\sum_l\frac{1}{\sqrt{2\pi}}\sum_{m,n} \ee^{\ii(m\phi_0-\omega_{mn} t_0)}
\sum_{i,s}
\MC_{si}(m,n,r_0)
\sum_{l_1,l_2}
\Psi_{l_1mn}^\pIH\R[\pIH,(i)]{2}{l_1mn}(r_0)\Sb{2}{mn}{l_1}{l_2}\YA{s}{m}{l_2}{l}\Y{}{lm}(0)\\
&=
\sum_{l} h_{uu}^{l,\pIH},
\end{align}
\end{widetext}
where in the second line we utilized that $\Y{}{l_4m}(z_0)$ vanishes on the equator for odd values of $l_4+m$, and consequently that 
\begin{equation}
\sum_{l_4} \YB{\hh{\frac{1-(-1)^{l_3+m}}{2}}}{m}{l_3}{l_4}\Y{}{l_4m}(0)=\Y{}{l_3m}(0).
\end{equation}
Equation \eqref{eq:huu5} thus gives the explicit expression for the $l$-modes of $h_{uu}$ that we were after. Following the above steps the  coefficients $\MC_{si}(m,n,r_0)$ can be computed explicitly, with their final form given in Table \ref{tab:huuc}.
\begin{table*}[tp]
\input{huutab.tex}
\caption{This table shows all the non-zero coefficients $\MC_{si}(m,n,r_0)$ of $h_{uu}$ in Eq. \eqref{eq:huu5}. Here $r_0$ is the radial coordinate at the particle worldline, and $\Delta_0$ is $\Delta$ evaluated at $r=r_0$. The $u^i$ are the tetrad components of the four-velocity of the particle.}\label{tab:huuc}
\end{table*}
The procedure in this section was purposely written more general than strictly necessary for the calculation of $h_{uu}$. In principle, the procedure described here can be applied to any quantity constructed from the metric and its derivatives, such as the self-force. This will just yield a different set of coefficients $\MC_{si}(m,n,r_0)$. For more general quantities, the expansion coefficients $\YB{j}{m}{l_3}{l_4}$ will not disappear from the final expression.

\subsection{Completion}\label{sec:comp}
Although the Weyl scalars $\psi_0$ and $\psi_4$ contain most of the information about linear metric perturbations, they cannot contain all information. The simplest counterexamples are perturbations of the background Kerr metric within the Kerr family. These necessarily have $\psi_0=\psi_4=0$, and can thus not be distinguished from the zero perturbation on the basis of $\psi_0$ and $\psi_4$. The CCK procedure can thus at best return a representative of the equivalence class of metric perturbations with the same $\psi_0$ and $\psi_4$. In general, we can write
\begin{equation}
h_{\mu\nu}^\mathrm{full} = h_{\mu\nu}^\mathrm{CCK} + h_{\mu\nu}^\mathrm{comp},
\end{equation}
where $h_{\mu\nu}^\mathrm{CCK}$ is the metric perturbation obtained by constructing $\psi_4$ from $h_{\mu\nu}^\mathrm{full}$ and then applying the CCK procedure. The procedure described in the previous sections will produce (the regular part of) $h_{\mu\nu}^\mathrm{CCK}$ produced by a particle moving on an equatorial orbit. However, to calculate $\Delta U$, we need  $h_{\mu\nu}^\mathrm{full}$; i.e. we need to find $h_{\mu\nu}^\mathrm{comp}$. This is known as the ``completion problem''.

The completion problem is made tractable by a very useful theorem due to Wald \cite{Wald:1973}. Wald showed that, up to gauge terms, the completion part in any vacuum region is given by,
\begin{equation}
h_{\mu\nu}^\mathrm{comp} =c_M h_{\mu\nu}^M + c_J h_{\mu\nu}^J +c_C h_{\mu\nu}^C +c_{NUT} h_{\mu\nu}^{NUT},
\end{equation}
where the metric perturbations $h_{\mu\nu}^*$ are obtained by embedding the background Kerr metric in the four-dimensional family of vacuum Pleba\'{n}ski-Demia\'{n}ski metrics \cite{Plebanski:1976gy} and varying with respect to the mass $M$, angular momentum $J=Ma$, C-metric acceleration $\alpha_C$, or NUT charge $q_{NUT}$, and the $c_*$ are constants.

This reduces the completion problem to finding four numbers in each of the two vacuum regions in our problem. The perturbations $h_{\mu\nu}^C$ and $h_{\mu\nu}^{NUT}$ have conical singularities extending from the horizon to infinity. Since the metric perturbation in the interior vacuum region must be regular at the horizon and the perturbation in the exterior vacuum region must be regular at infinity, we must have $c_C^\pIH=c_{NUT}^\pIH=0$.

Requiring that the perturbed spacetime has the correct ADM mass and angular momentum fixes the parameters in the exterior vacuum region,  $c_M^\pI=m\nE$ and $c_J^\pI=m\nL$. We obtain
\begin{equation}
\begin{split}
h_{\mu\nu}^{\mathrm{comp},\pI} = 2\mr \Bh{
\frac{\nE}{r}\id{t}^2
+\frac{r^2\hh{(r+a^2)\nE-a\nL}}{\Delta^2}\id{r}^2&
\\
+\frac{a\hh{(r+2)\nL-2a(r+1)\nE}}{r}\id{\phi}^2
-\frac{2\nL}{r}\id{t}\id\phi
}.&
\end{split}
\end{equation}
By requiring continuity of certain gauge invariant quantities away from the equatorial plane, Merlin et al.  \cite{Merlin:2015} have shown that for particles moving on equatorial orbits $c_M^\pH=c_J^\pH=0$.

Wald's theorem only fixes the completion part up to gauge modes. Recall that to calculate $U$ we must be in a gauge that is asymptotically flat and respects the periodic structure of the orbit.  The two components of the metric perturbation in the exterior vacuum region, $h_{\mu\nu}^{\mathrm{CCK},\pI}$ and $h_{\mu\nu}^{\mathrm{comp},\pI}$, naturally satisfy these conditions. On the other hand, the reconstructed metric perturbation in the interior vacuum region $h_{\mu\nu}^{\mathrm{CCK},\pH}$ has (gauge) string like singularity extending towards infinity \cite{Pound:2013faa}. We must therefore construct $\Delta U$ from the values of $h_{uu}$ obtained from the exterior vacuum region.

\subsection{Regularization}\label{sec:reg}
The procedure above calculates $h^\ret_{uu}$ constructed from the retarded solution to the linearized Einstein equation. This quantity obviously diverges at the location of the particle. This manifests itself in \eqref{eq:huu5} as the outer sum over $l$ diverging, while the individual $l$-modes, $h_{uu}^l$ are all finite.

A key result of the self-force program is that the retard metric perturbation can be split into a smooth regular ``$\reg$'' piece, and a singular ``$\sing$'' piece, in such a way that the contribution of the singular ``$\sing$'' piece to the equations of motion of the particle vanish (see \cite{Poisson:2011nh} and the references therein). Moreover, this split can be arranged in such a way that the regular ``$\reg$'' piece is a vacuum solution to the linearized Einstein equation and the particle follows a geodesic effective space time
\begin{equation}
\tilde{g}_{\mu\nu} = g_{\mu\nu} + h_{\mu\nu}^\reg.
\end{equation}
Consequently, as mentioned in Sec. \ref{sec:GSF}, $\Delta U$ is defined with respect to this regular effective spacetime. Hence we want to calculate
\begin{equation}
h_{uu}^\reg(x_0(\mt)) =\lim_{x\to x_0(\mt)} \bh{h_{uu}^\ret(x)-h_{uu}^\sing(x)}.
\end{equation}
Both $h_{uu}^\ret$ and $h_{uu}^\sing$ diverge at the particle. Since the individual $l$-mode contributions to $h_{uu}^\ret$ and $h_{uu}^\sing$ are finite at the particle one can write following \cite{Barack:2001gx},
\begin{align}
h_{uu}^\reg(x_0) &= \lim_{x\to x_0} \bh{h_{uu}^\ret(x)-h_{uu}^\sing(x)}
\\
&= \sum_{l=0}^\infty \hh{h_{uu}^{\ret,l}(x_0)-h_{uu}^{\sing,l}(x_0)}.
\end{align}
A key idea behind the mode-sum regularization scheme is that the $l$-modes of the singular field can be written as
\begin{equation}
h_{uu}^{\sing,l} = \RP{0} +\frac{\RP{1}}{l+\frac{1}{2}} +\bigO\bh{\frac{1}{(l+\frac{1}{2})^2}}.
\end{equation}

In general this expansion depends on the chosen gauge and extension of $u^\mu$ and more generally the extension of the individual terms of $h_{uu}$. It is most easily calculated in the Lorenz gauge. Working in that gauge (and the same ``rigid'' $u^\mu$ is constant extension as used in Sec. \ref{sec:MR}), one can show \cite{Heffernan:2012vj} that for a particle moving on an equatorial orbit around a Kerr black hole,
\begin{align}
(B = )\quad& 
\RP{0}^\Lor
 =\frac{
4 K(\frac{
\nL^2+a^2 +\frac{2a^2}{r_0}
}{
\nL^2+r_0^2+a^2 +\frac{2a^2}{r_0}
})
}{
\pi\sqrt{\nL^2+r_0^2+a^2 +\frac{2a^2}{r_0}}
} \label{eq:B},
\\
(C =)\quad& \RP{1}^\Lor =0,\\
(D=)\quad&\sum_{l=0}^\infty h_{uu}^{\sing,l} - \RP{0} -\frac{\RP{1}}{l+\frac{1}{2}}= 0 ,
\end{align}
where the letters in parentheses are the traditional names for these ``regularization parameters'' used in \cite{Barack:2001gx,Barack:2002bt,Barack:2009ux}. Vanishing of the $C$ and $D$ parameters means that the mode-sum for the regular part of $h_{uu}$ simply becomes (in Lorenz gauge),
\begin{equation}
h_{uu}^\reg = \sum_{l=0}^{\infty}\hh{h_{uu}^{\ret,l}-\RP{0}}.
\end{equation}
However, the metric reconstructed using the CCK procedure is obtained in an outgoing radiation gauge. 
A main obstruction in radiation gauge is that in general the full retarded metric perturbation is also singular away from the particle with a string like singularity radiating from the particle in one of the principle null directions. 

An in depth analysis of this problem was given by Pound et al. in \cite{Pound:2013faa}. They calculate the radiation gauge regularization parameters of the self-force by finding a local gauge transformation that transforms the radiation gauge to a gauge that locally near the particle resembles the Lorenz gauge. This ``locally Lorenz gauge'' falls in the class of gauges related to Lorenz gauge by a continuous gauge transformation, for which one can show that one can use the Lorenz form of the mode-sum formula. Pound et al. find that for self-force in one of the ``half-string'' radiation gauges (regular in either the interior or exterior region) the $A$, $B$, and $C$ parameters are unchanged, but the $D$ parameter gains a finite correction.

Using the results from \cite{Pound:2013faa} it is elementary to extend their analysis to the regularization parameters for $\avg{h_{uu}}$ in radiation gauge. Let $\xi_\mu$ be the gauge vector that transforms the radiation gauge to a locally Lorenz gauge. Under its action the change in $h_{uu}$ is
\begin{align}
\delta_\xi h_{uu} &= u^\mu u^\nu \delta_\xi h_{\mu\nu}
\\
&=  2 u^\mu u^\nu \CD{\mu}\xi_\nu
\\
&=  2 u^\mu\CD{\mu}(u^\nu\xi_\nu)- 2\xi_\nu u^\mu\CD{\mu}u^\nu.\label{eq:deltahuu}
\end{align}
The first term is a proper time derivative of a spacetime scalar. Consequently, it will vanish after averaging over the orbit. In \cite{Pound:2013faa} it is shown that
\begin{align}
\xi_\mu &=\bigO(\log{s}),\text{and}\\
u^\mu\CD{\mu}u^\nu &= \bigO(s),
\end{align}
where $s$ is the separation from the worldline. Consequently, the second term in \eqref{eq:deltahuu} vanishes as one takes the limit to the worldline. We thus find that $\avg{h_{uu}}$ is in fact invariant under the gauge transformation defined by $\xi_\mu$. Therefore, we find that the correction to the regularization parameters for $\avg{h_{uu}}$ in radiation gauge must be zero.

The mode-sum formula $\avg{h_{uu}}$ in radiation gauge is thus,
\begin{equation}
\avg{h_{uu}^{\reg,\Rad,\pIH}} = \sum_{l=0}^{\infty}\hh{\avg{h_{uu}^{\ret,\Rad,l,\pIH}}-\avg{\RP{0}^\Lor}}.
\end{equation}
We can apply this formula to numerically obtain the regular field from our reconstructed metric.

\section{Numerical Implementation}\label{sec:implementation}
We have developed a numerical code in \emph{Mathematica} which implements the above method for a mass following an equatorial geodesic around a Kerr black hole. The code uses the following steps:
\begin{enumerate}
\item Calculate the geodesic sourcing the gravitational perturbation. (To be discussed in Sec. \ref{sec:geodesics}.)
\item Solve the Teukolsky equation to obtain the linear perturbation to $\psi_4$. (To be discussed in Sec. \ref{sec:MST}.)
\item Reconstruct the linear (retarded) metric perturbation and $h_{uu}^\pIH$.
\item Regularize $h_{uu}^\pIH$ using the mode-sum formula. (See Sec. \ref{sec:reg}.)
\item Estimate the contribution from the high $l$ tail. (To be discussed in Sec. \ref{sec:tailest}.)
\item Add the contribution from the completion. (See Sec. \ref{sec:comp}.)
\end{enumerate}
Step 1 only needs to be done once, and is done in a matter of seconds. Steps 2 and 3 need to be iterated over all modes. Section \ref{sec:truncation} discusses the criteria used to truncate the infinite sums over modes. Once this loop is complete the remaining steps can be done in one go.

\subsection{Geodesics}\label{sec:geodesics}
We start our calculation by specifying a spin $a$ for the central black hole, and an eccentricity $e$ and semilatusrectum $p$, which define an equatorial bound geodesic orbit. From $a$, $p$, and $e$ we calculate the values of the (specific) energy ($\nE$), (specific) angular momentum ($\nL$), orbital frequencies ($\Omega_r$ and $\Omega_\phi$), and periods ($\ctrp$, $\ptrp$, and $\mtrp$) using the analytical formulas from \cite{Schmidt:2002qk,Fujita:2009bp}. 

A key observation used throughout the code is that all the quantities that we are interested in are smooth periodic functions along the orbit. As a consequence, numerical calculations of orbital integrals and averages converge exponentially with the number of sampling points, if one uses an even sampling. To make maximal use of this, we fix an evenly spaced (in Mino time) grid of $N$ points along the orbit, and precomputed as many quantities along the orbit as possible. The number of points $N$ will determine the precision of the orbital integrals used to obtain the orbital averages and the coefficients $Z_{lmn}^\pIH$ of the inhomogeneous solutions of the Teukolsky equation. In practice $n$ will range from 1 for circular orbits to about 400 for the most eccentric orbits ($e=0.4$) that we calculate here.

We start by computing the geodesics. Fujita and Hikida \cite{Fujita:2009bp} have derived analytic solutions for bound geodesics around a Kerr black hole in terms of elliptic functions. Evaluation of elliptic functions is relatively costly. To save computation time we evaluate the analytic solutions (and their first derivatives) on our orbital grid and save these as arrays for future use.

Further savings can be made by precomputing some functions along the orbit that will be used frequently. For example,  the coefficients in Table \ref{tab:huuc} can be split in terms independent of $l$, $m$, and $n$, which are functions along the orbit that can be computed along the orbit. To avoid recalculating these functions for each mode, we evaluate them once on the grid at this stage of the computation and cache their values. A similar  procedure can be used for the coefficients appearing in the source term $\T{-2}{lmn}(r)$ in \eqref{eq:sourceint} (see \cite{Meent:2015a}).

\subsection{MST method}\label{sec:MST}
We solve the radial Teukolsky equation using the semi-analytical method developed by Mano, Suzuki, and Takasugi (MST) \cite{Mano:1996vt,Mano:1996gn}, see \cite{ST:lrr-2003-6} for a review. In this method, solutions to the homogeneous radial Teukolsky equation are obtained as a series expansion in certain hypergeometric functions. Details of our numerical implementation can be found in \cite{Meent:2015a}, which refines numerical methods developed in \cite{Fujita:2004rb,Fujita:2009us,Throwethesis}.

When presented with the orbital data calculated in the first step, the code detailed in \cite{Meent:2015a} will return the homogeneous solutions $\R[\pIH]{-2}{lmn}$ and their first radial derivatives evaluated at the grid points along the orbit. In addition the code returns values for the coefficients of the inhomogeneous solutions $Z^\pIH_{lmn}$, and the coefficients of the asymptotic expansion at infinity and the horizon of the homogeneous solutions, $\A[\pIH]{-2}{lmn}$, $\B[\pIH]{-2}{lmn}$, $\Ainf[\pIH]{-2}{lmn}$, and $\Binf[\pIH]{-2}{lmn}$.

We can then use Eqs. \eqref{eq:psiI},\eqref{eq:psiH},\eqref{eq:R2I}, and \eqref{eq:R2H} to construct the value of the modes of the Hertz potential at each grid point along the orbit. The first radial derivatives are obtained from the radial derivatives of Eqs. \eqref{eq:R2I} and \eqref{eq:R2H}. Any higher order radial derivatives ($h_{uu}$ needs only the second radial derivative) can be obtained from recurrence relations obtained from the radial Teukolsky equation (see \cite{Meent:2015a}). 

\subsection{Truncation}\label{sec:truncation}
We now have all ingredients needed to evaluate Eq. \eqref{eq:huu5}. However, we need to decide how to truncate the infinite sums (over, $l$, $l_i$, $n$ and $m$). Since the (retarded) metric perturbation is divergent, we do not expect the outer sum over $l$ to converge. Instead the ``$l$-modes'',
\begin{equation}\label{eq:lmode}
\begin{split}
&h_{uu}^{l,\pIH} =\frac{1}{\sqrt{2\pi}}\sum_{m,n} \ee^{\ii(m\phi_0-\omega_{mn} t_0)}
\sum_{i,s}
\MC_{si}(m,n,r_0)
\\
&\quad\times
\sum_{l_1,l_2}
\Psi_{l_1mn}^\pIH\R[\pIH,(i)]{2}{l_1mn}(r_0)\Sb{2}{mn}{l_1}{l_2}\YA{s}{m}{l_2}{l}\Y{}{lm}(0),
\end{split}
\end{equation}
are expected to converge to a constant value $\RP{0}$ given by \eqref{eq:B}. Note that a single $lmn$ mode of $h_{uu}$ requires data from $l_1mn$ modes of the Hertz potential with the same $m$ and $n$ but different various values of $l_1$. This suggests that the most natural way to loop over the modes is to first fix values for $m$ and $n$ and calculate all $l_1mn$ modes of the Hertz potential up till a certain predetermined $\llmax$. If we treat the $\Psi_{l_1mn}^\pIH\R[\pIH,(i)]{2}{l_1mn}(r_0)$'s with different $l_1$ as forming a vector, then
\begin{equation}
\begin{split}
M_{l_1,l} = \frac{1}{\sqrt{2\pi}}\ee^{\ii(m\phi_0-\omega_{mn} t_0)}&\MC_{is}(m,n,r_0)
\\
&\times\sum_{l_2}\Sb{2}{mn}{l_1}{l_2}\YA{s}{m}{l_2}{l}
\end{split}
\end{equation}
defines a linear operator that produces a vector of contributions to the $h^{l,^\pIH}_{uu}$'s. The re-expansion of the spin-weighted spherical harmonics to ordinary spherical harmonics, $\YA{s}{m}{l_2}{l}$, only introduces a coupling between a finite number of $l_1$-modes. The re-expansion for spheroidal to spherical harmonics $\Sb{2}{mn}{l_1}{l_2}$, on the other hand, causes a coupling between an infinite number of $l_1$-modes. However, $M_{l_1,l}$ can be seen to decay exponentially away from the diagonal. We can thus control the error caused by the finite truncation by choosing a maximum value $\lmax$ for $l$ that is smaller than $\llmax$. In practice, the code (at each contribution to an $mn$-mode) determines the largest $\lmax$ such that
\begin{equation}
\max_{l\leq \lmax}\hh{ \abs{\frac{
M_{\llmax+1,l}
}{
M_{l,l}
}} +
\abs{\frac{
M_{\llmax+2,l}
}{
M_{l,l}
}} }< \TP.
\end{equation}
The final $\lmax$ will be the minumum of the $\lmax$'s determined at each mode. We can now sum the contributions of all $i$, $s$, and $j$ in \eqref{eq:lmode} to obtain all $lmn$-modes of $h_{uu}$ with fixed $m$ and $n$ up to $\lmax$,
\begin{equation}\label{eq:lmnmode}
\begin{split}
&h^{lmn,\pIH}_{uu} = \frac{1}{\sqrt{2\pi}}\Bh{ \ee^{\ii(m\phi_0-\omega_{mn} t_0)}\sum_{i,s,j}\MC_{is}(m,n,r_0)
\\
&\hspace{8pt}\times\sum_{l_1,l_2}\Psi_{l_1mn}^\pIH\R[\pIH,(i)]{2}{l_1mn}(r_0)\Sb{2}{mn}{l_1}{l_2}\YA{s}{m}{l_2}{l}}\Y{}{lm}(0).
\end{split}
\end{equation}
For the final sum over $m$ and $n$ we observe that
\begin{equation}
h_{uu}^{lmn,\pIH} = h_{uu}^{l{-m}{-n},\pIH},
\end{equation}
and use this to limit our computation to modes with $n\geq 0$, obtaining the rest by symmetry. Furthermore, to calculate $\Delta U$ we only need the orbital average $\avg{h_{uu}^{lmn}}$. For fixed $l$ and $m$, and large enough $n$, the magnitude of $\avg{h_{uu}^{lmn,\pIH}}$ decays exponentially with increasing $n$. This leads us to the following truncation procedure. Starting at $n=0$, we calculate  $\avg{h_{uu}^{lmn,\pIH}}$ for all $\abs{m}\leq \lmax$.  For each next $n$ we calculate $\avg{h_{uu}^{lmn,\pIH}}$, for the same set of $m$ as $n-1$, except for those $m$'s for which
\begin{equation}
\max_{l\leq \lmax} \abs{\frac{\avg{h_{uu}^{lmn-1,\pIH}}}{\avg{\RP{0}^\Lor}}} < \TP.
\end{equation}
In addition, each computation of $\avg{h_{uu}^{lmn,\pIH}}$ will also produce an estimate of the relative error on its value. Generally, this error will increase as we increase $n$, mainly because the integrand \eqref{eq:sourceint} becomes more oscillatory. If for some $n$ the maximum estimated relative error for $\avg{h_{uu}^{lmn,\pIH}}$ becomes larger than our target precision $\TP$, we adjust our target precision upwards to match the estimated error.

We continue this procedure until $\avg{h_{uu}^{lmn,\pIH}}$ has converged to our target precision $\TP$ for all $m$.

\begin{table*}[t]
\begin{tabular}{d{-1}d{-1}|d{20}d{14}d{14}}
 \hline\hline
 p		& e	   & \TabH{Here}	& \TabH{Akcay et al. \cite{Akcay:2015pza}}	& \TabH{Barack and Sago \cite{Barack:2011ed}} \\
 \hline
 10 	& 0.10 & -0.1277540232(10) 		& -0.1277540(3)		& -0.1277554(7)\\
 15 	& 0.10 & -0.07687063237(5) 		& -0.0768706(2)		& -0.0768709(1)\\
 20 	& 0.10 & -0.055221659739(6)		& -0.05522166(7)	& -0.05522177(4)\\
 100 	& 0.10 & -0.010101234326660(2) 	& -0.0101012344(10)	& \text{---}\\
\hline
 10		& 0.20 & -0.123647888(2) 			& -0.123648(3)		& -0.1236493(7)\\
 15 	& 0.20 & -0.07431375582(4) 		& -0.07431376(9)	& -0.0743140(1)\\
 20 	& 0.20 & -0.0534085449572(6) 		& -0.05340854(9)	& -0.05340866(4)\\
 100 	& 0.20 & -0.0097893279221005(14) 	& -0.0097893274(4)	& \text{---}\\
\hline
 10 	& 0.30 & -0.1168019818(2) 		& -0.1168020(6)		& -0.1168034(6)\\
 15 	& 0.30 & -0.07007684538(10) 		& -0.0700768(5)		& -0.0700771(1)\\
 20 	& 0.30 & -0.050403774160(6) 		& -0.05040377(4)	& -0.05040388(4)\\
 100 	& 0.30 & -0.009270280959(2) 		& -0.009270281(4)	& \text{---}\\
\hline
 10 	& 0.40 & -0.107220(6) 			& -0.107221(2)		& -0.1072221(5)\\
 15 	& 0.40 & -0.0641988(4) 			& -0.064199(1)		& -0.0641991(1)\\
 20 	& 0.40 & -0.0462337(4) 			& -0.0462337(9)		& -0.04623383(4)\\
 100 	& 0.40 & -0.0085452(4) 			& -0.0085453(2)		& \text{---}\\
  \hline\hline
\end{tabular}
\caption{Numerical results for $\Delta U$ from our new code for eccentric orbits around a Schwarzschild $a=0$ compared with previous frequency domain from  Akcay et al. \cite{Akcay:2015pza} and time domain results from Barack and Sago \cite{Barack:2011ed}.}\label{tab:SSecc}
\end{table*}
\subsection{Tail estimation}\label{sec:tailest}
Adding together all $m$ and $n$ modes, the procedure above will give us all $\avg{h_{uu}^{l,\pIH}}$ up to $\lmax$. This allows us to get an estimate of $\avg{h^{\reg,\pIH}_{uu}}$,
\begin{equation}
\avg{h^{\reg,\pIH}_{uu}} \approx \sum_{l=0}^{\lmax}\bh{ \avg{h_{uu}^{l,\pIH}}-\avg{\RP{0}}}.
\end{equation}
However, we expect $\avg{h_{uu,\pIH}^{l}}-\avg{\RP{0}^\Lor}$ to be order $l^{-2}$. Consequently, the missing tail,
\begin{equation}
\sum_{l=\lmax+1}^{\infty}\bh{ \avg{h_{uu}^{l,\pIH}}-\avg{\RP{0}^\Lor}} \sim \frac{1}{\lmax}.
\end{equation}
With a typical value of $\lmax=20$ this would gives us only two digits of precision. We know the individual $l$-modes to a much higher precision. We can leverage that fact to get an estimate of the missing tail.

Our procedure is as follows. We first calculate the partial sums
\begin{equation}
S_L^\pIH \equiv  \sum_{l=0}^{L} \bh{\avg{h_{uu}^{l,\pIH}}-\avg{\RP{0}^\Lor}}.
\end{equation}
We can model $S_L$ as an expansion in $1/L$
\begin{equation}\label{eq:Smodel}
S_L^\pIH = \sum_{k=0}^{\infty} \frac{\mathcal{S}_k^\pIH}{L^k},
\end{equation}
with the total sum being estimated as
\begin{equation}
\avg{h^{\reg,\pIH}_{uu}} =\lim_{L\to\infty} S_L^\pIH = \mathcal{S}_0^\pIH.
\end{equation}
We can estimate $\mathcal{S}_0^\pIH$ by fitting a truncated version of the model \eqref{eq:Smodel},
\begin{equation}
\sum_{k=0}^{\kmax} \frac{\mathcal{S}_k^\pIH}{L^k},
\end{equation}
to the $S_L^\pIH$. This fit is improved by weighting the data point $S_L^\pIH$ with their estimated errors. In practice (especially for low $L$) this error is dominated by the part of \eqref{eq:Smodel} that is not included in the truncated model. If we assume that all $\mathcal{S}_k^\pIH$ are order unity, we can estimate this error to be approximately $L^{-\kmax-1}$. This estimate is good for all but the smallest $L$, consequently we want to exclude those points from the fit. Similarly, the data points with the largest $L$ may be so dominated by the truncation errors that they would introduce a systematic bias in the fit. We thus fit the $S_L$ data in a window $\Lmin\leq L\leq \Lmax$.

To find the optimal values for $\Lmin$, $\Lmax$, and $\kmax$, we perform the fit for a wide range of (reasonable) choices for their values. For each fit we calculate the estimated error on $\mathcal{S}_0^\pIH$. From these fits we choose the ones with the lowest estimated error on $\mathcal{S}_0^\pIH$. We then estimate $\mathcal{S}_0^\pIH$ to be the median prediction from this set, and estimate the error in  $\mathcal{S}_0^\pIH$ to be the maximum of the error estimated from the fit and the median deviation of the estimates of $\mathcal{S}_0^\pIH$ among the best fits.

This allows us to get a much more accurate estimate of $\avg{h^{\reg,\pIH}_{uu}}$. In particular, this estimate can be more accurate than the estimated errors on the $l$-modes with the largest $l$.

\section{Results}\label{sec:results}

\subsection{Comparison with existing results}
\begin{table}
\begin{tabular}{r|rrr}
\hline\hline
Term	&\multicolumn{3}{c}{Coefficient}\\
	&	\TabH{Est. Here}	& \TabH{Est. \cite{Akcay:2015pza}}	& Exact PN\cite{Akcay:2015pza}\\
\hline
$\frac{e^2}{p^2}$	& $4\pm 6\cdot 10^{-12}$				& $4.0002(8)$				& $4$\\
$\frac{e^4}{p^2}$	& $-2\pm 4\cdot 10^{-10}$	& $-2.00(1)\phantom{00}$	& $-2$\\
$\frac{e^6}{p^2}$	& $0\pm 4\cdot 10^{-9\phantom{0}} $	& 							& 0\\
\hline
$\frac{e^2}{p^3}$	& $7\pm 6\cdot 10^{-9}$	& $7.02(3)\phantom{00}$		& $7$\\
$\frac{e^4}{p^3}$	& $\frac{1}{4} \pm 4\cdot 10^{-7}$	&							& $\frac{1}{4}$\\
$\frac{e^6}{p^3}$	& $\frac{5}{2} \pm 4\cdot 10^{-6}$	& 							& $\frac{5}{2}$\\
\hline
$\frac{e^2}{p^4}$	& $-14.3120980(5)$	& $-14.5(4)\phantom{000}$	& $\approx -14.31209731$\\
$\frac{e^4}{p^4}$	& $83.38298(7)\phantom{00}$	& 							& $\approx\phantom{-} 83.38296351$\\
$\frac{e^6}{p^4}$	& $-36.421(3)\phantom{0000}$	& 							& $\approx -36.42197567$\\
\hline
$\frac{e^2}{p^5}$			& $-345.37(5)\phantom{00000}$	& $-1500\pm 500\phantom{.0}$&  \\
$\log{p}\frac{e^2}{p^5}$	& $19.733(5)\phantom{0000}$		& $250\pm 	100\phantom{.0}$&  \\
$\frac{e^4}{p^5}$			& $737(4)\phantom{0.000000}$	& 							&  \\
$\log{p}\frac{e^4}{p^5}$	& $-48.6(4)\phantom{000000}$	& 							&  \\
$\frac{e^6}{p^5}$			& $-185(12)\phantom{0.00000}$	& 							&  \\
$\log{p}\frac{e^6}{p^5}$	& $7(2)\phantom{0.000000}$		& 							&  \\
\hline
$\frac{e^2}{p^6}$			&  $-2000\pm 400\phantom{0000}$	& 						&  \\
$\log{p}\frac{e^2}{p^6}$	&  $-40\pm 20\phantom{00000}$	& 						&  \\
\hline\hline
\end{tabular}
\caption{Numerical estimates of the coefficients of the expansion of $\Delta U$ in Schwarzschild as a power series in $p^{-1}$ and $e$. The first column shows numerical estimates obtained by fitting a large data set of eccentric orbits in Schwarzschild with $10\leq p\leq1000$ and $0\leq e\leq 0.4$ obtained using our new code. The second column shows the numerical results from Akcay at el. \citep{Akcay:2015pza}. The last column show the exactly known values of the coefficients up to 3PN.}\label{tab:PN}
\end{table}
\begin{table*}[tbp]
\input{circkerrtab.tex}
\caption{Numerical results for $\Delta U$ for circular orbits in Kerr from our new code, compared with previous results from the circular orbit code of Shah et al. \cite{Shah:2012gu}. In each cell, the top number in each cell is the result from our new code, and the bottom number the result from \cite{Shah:2012gu}. (Note that these are not the values published in \cite{Shah:2012gu}, but values from an updated version of that code with some errors fixed.)  Parenthetical figures indicate the estimated error on the last displayed digit. When no parenthetical figure the error is at least an order of magnitude smaller than the last shown digit.}
\label{tab:kerrcirc}
\end{table*}
\begin{figure*}[tp]
\begin{minipage}{.48\textwidth}
\includegraphics[width=\columnwidth]{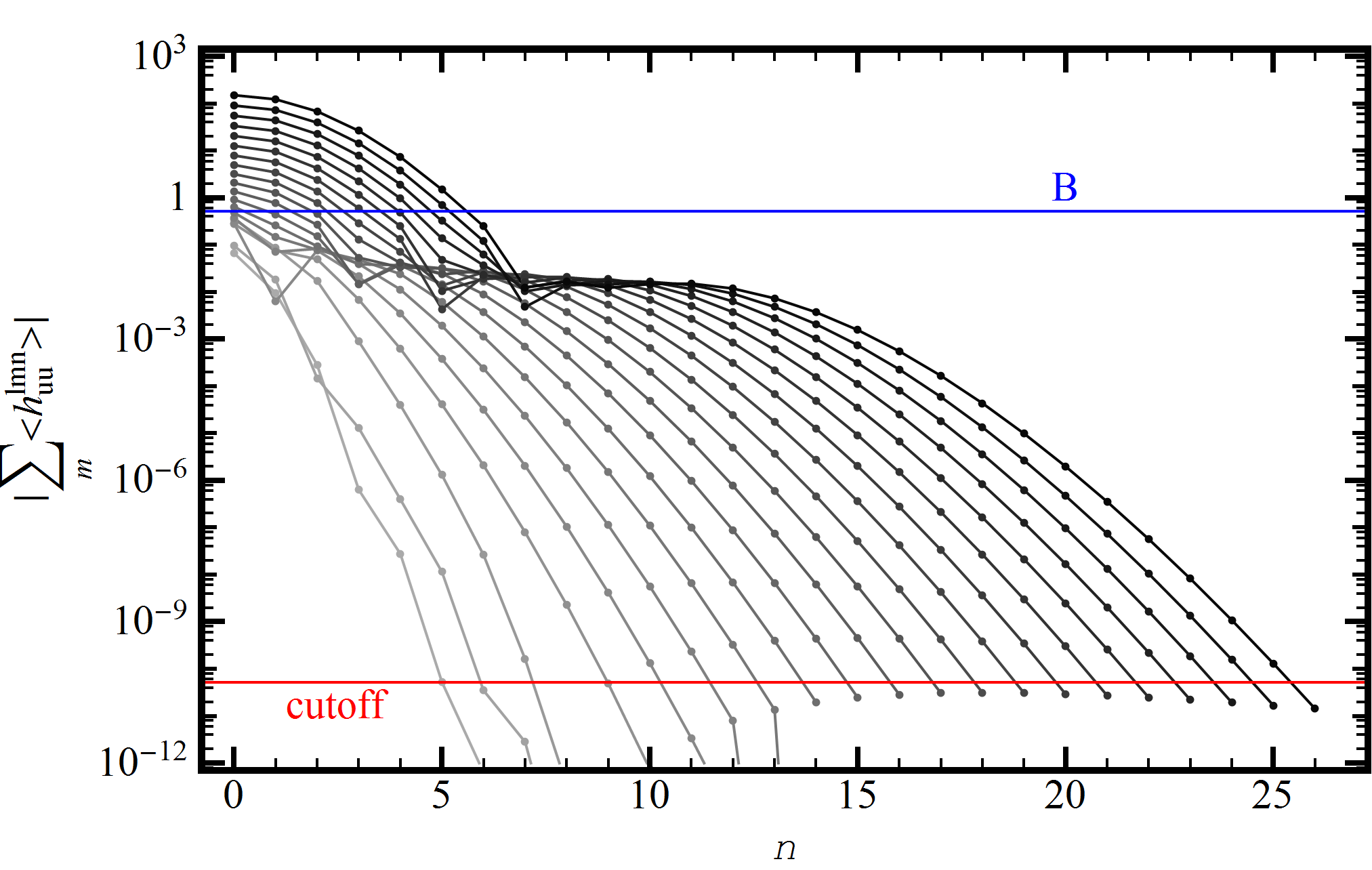}
\caption{Convergence of the $l$-modes with increasing $n$-modes (for an orbit with spin $a=0.9$, semilatus rectum $p=3.32$, and eccentricity $e=0.2$). Each line represents the $n$-mode contributions to one $l$-mode with darkers shades representing higher values of $l$. The two horizontal lines represent the target accuracy and the regularization parameter $B$ to which the sum of the $n$-modes will (approximately) converge for each $l$-mode.}\label{fig:lnconvplot}
\end{minipage}\hspace{.04\textwidth}%
\begin{minipage}{.48\textwidth}
\includegraphics[width=\columnwidth]{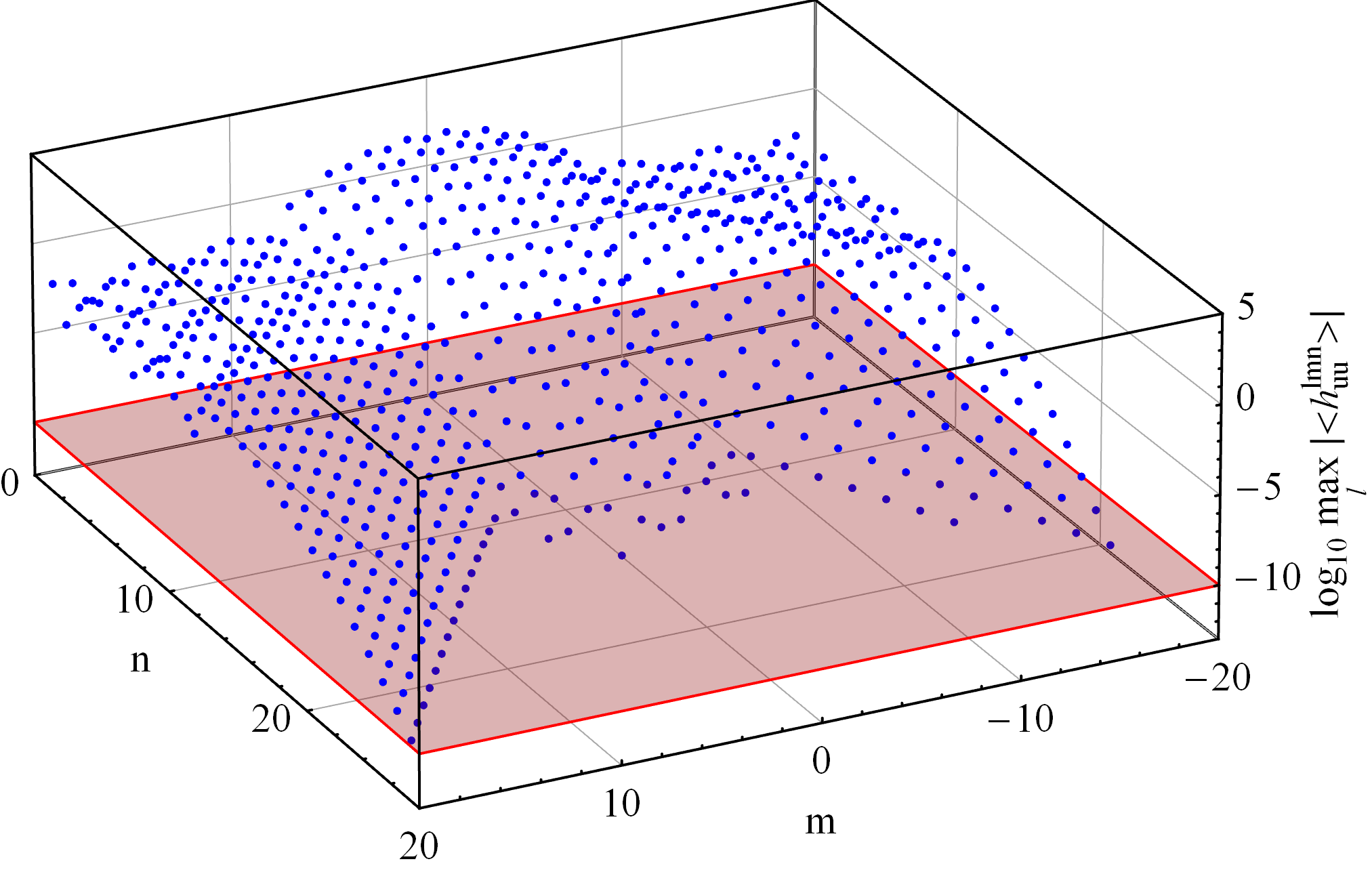}
\caption{The maximal contribution to the $l$-modes at each $m$ and $n$  $n$-modes (for an orbit with spin $a=0.9$, semilatus rectum $p=3.32$, and eccentricity $e=0.2$). The horizontal plane represents the value of the accuracy goal.}\label{fig:mnconvplot}
\end{minipage}
\end{figure*}
An important verification of the numerical procedure discussed in this paper is that it reproduces existing results from the literature in the relevant limits. The generalized redshift invariant for eccentric orbits around a non-spinning Schwarzschild black hole was first calculated by Barack and Sago \cite{Barack:2011ed} using a time domain based code. Later, Akcay et al. \cite{Akcay:2015pza} published more accurate results using their frequency domain code.

In table \ref{tab:SSecc}, we compare results from our new code with these previous publications. For low eccentricities our results are consistently within the error bars from \cite{Akcay:2015pza} providing more significant digits. In fact, our more accurate results suggest that the error bars given in \cite{Akcay:2015pza} are slightly pessimistic. Whereas the results from \cite{Akcay:2015pza} and \cite{Barack:2011ed} were marginally compatible due to overlapping error bars, the tighter error bars on the new data are inconsistent with the last digit of \cite{Barack:2011ed}. It appears the time domain results consistently overestimate the magnitude of $\Delta U$.

The results for \cite{Akcay:2015pza} were generated using the same computer cluster in Southampton. This gives us the opportunity to compare the computational efficiency of our new semi-analytical \emph{Mathematica} based code, with the older C-based frequency domain code which numerically solved the linearized field equations. Running on two 16 processor nodes, the old code took 3915 seconds to calculate all the modes for the orbit with $p=10$ and $e=0.1$. Running on just one node our code took just 1009 seconds to reproduce their result at a similar precision.

In \cite{Akcay:2015pza}, a procedure for extracting the coefficients of an expansion of $\Delta U$ in powers of $e$ and $p$ from their data was described, comparing the result to the post-Newtonian (PN) predictions. It is relatively easy to produce a similar dataset with our new code, but at a higher accuracy, and follow their procedure. The results are listed in table \ref{tab:PN}. We were able to extract all known PN coefficients up to $e^6$ accurate to at least five digits. 

We also obtain fairly accurate estimates for the $e^2$ and $e^4$ 4PN coefficients, which will also have $\log p$ terms. The $e^2$ coefficients probably have a similar expansion in transcendental numbers as the 4PN terms of the expansion of $\Delta U$ for circular orbits. This means that the coefficient of $e^2 p^{-5}$ probably is some rational combination of $\log 2$, $\pi^2$, and the Euler-Mascheroni constant $\gamma$. With the limit number of digits available we are unable to guess what this combination is. The coefficient of $e^2 p^{-5}\log{p}$, on other hand is probably a rational number. Based on our numerical fit, it is most likely $99/5$. We finally get some very rough estimates for the coefficients of the $e^6$ 4PN and $e^2$ 5PN terms, which at least get the sign and order of magnitude correctly.

Shah et al. \cite{Shah:2012gu} previously calculated the redshift invariant for a particle in a circular equatorial orbit around a Kerr black hole using similar techniques as here, but numerically solving the Teukolksy equation. In Table \ref{tab:kerrcirc} we compare their results with the results from our new code. The results agree to the expected accuracy. In general the new code matches the accuracy of the code from \cite{Shah:2012gu} with exception of some of the strong field modes. The accuracy is currently limited by the highest value of $\llmax=30$ that our MST Teukolsky solver \cite{Meent:2015a} can currently handle, whereas \cite{Shah:2012gu} goes up to $\llmax=80$. There is no principle obstruction to raising the $\llmax$ limit of our code, it simply requires some time and effort to further populate some lookup tables used to seed the numerical root finding routines.

\subsection{Generalized redshift for eccentric orbits in Kerr}
\begin{figure}[tpb]
\includegraphics[width=\linewidth]{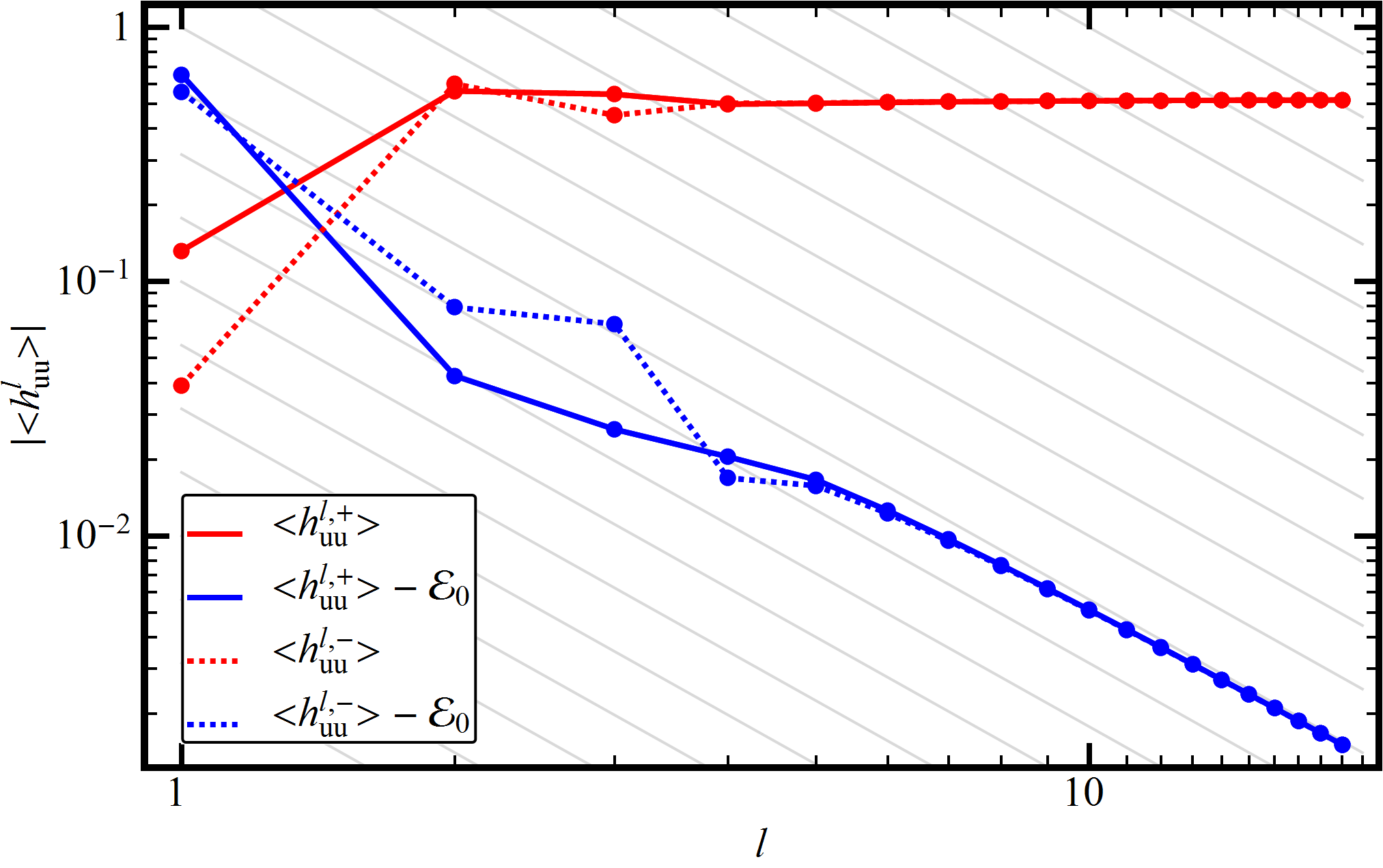}
\caption{The l-modes $h_{uu}^{l,\pIH}$ for an orbit with spin $a=0.9$, semilatus rectum $p=3.32$, and eccentricity $e=0.2$. The top lines are the values of the retard field before subtraction of the regularization parameters. The bottom lines give the values after subtraction of the regularization parameters. They follow the diagonal gridlines which represent $l^{-2}$ decay.}\label{fig:lmodeplot}
\end{figure}
\begin{table*}[tp]
\begin{tabular}{ld{-1}|d{18}d{18}d{18}}
\hline\hline
 p	&	e	&\multicolumn{1}{c}{$a=-0.9$}	& 	\multicolumn{1}{c}{$a=0.0$}	& \multicolumn{1}{c}{$a=0.9$}\\
\hline
 $\pISO+1$ & 0.00 & -0.1591225238(2) & -0.2208475274(2) & -0.40871659(2) \\
 $\pISO+1$ & 0.10 & -0.150998651(3) & -0.209375588(9) & -0.391205(6) \\
 $\pISO+1$ & 0.20 & -0.142034175(2) & -0.195877797(5) & -0.3635027(6) \\
 $\pISO+1$ & 0.30 & -0.13127905(3) & -0.179591180(7) & -0.327109(4) \\
 $\pISO+1$ & 0.40 & -0.1182790(4) & -0.160212(3) & -0.283764(2) \\
 \hline
 $\pISO+10$ & 0.00 & -0.062991300291(4) & -0.072055057429096(4) & -0.09093033701(3) \\
 $\pISO+10$ & 0.10 & -0.061183242(1) & -0.070241089(2) & -0.089218442(2) \\
 $\pISO+10$ & 0.20 & -0.058180786(3) & -0.066951046(3) & -0.085406440(1) \\
 $\pISO+10$ & 0.30 & -0.054043290(2) & -0.06226898(1) & -0.0796315937(2) \\
 $\pISO+10$ & 0.40 & -0.048833176(5) & -0.0562888674(2) & -0.072058378(4) \\
 \hline
 $\pISO+100$ & 0.00 & -0.009395362239424(0) & -0.009616383265554(0) & -0.009942932464012(0) \\
 $\pISO+100$ & 0.10 & -0.00927472002(8) & -0.00950017309(4) & -0.0098334067(4) \\
 $\pISO+100$ & 0.20 & -0.0089653010(1) & -0.00918951172(9) & -0.0095209535(1) \\
 $\pISO+100$ & 0.30 & -0.00846904194(5) & -0.00868616413(8) & -0.00900719351(3) \\
 $\pISO+100$ & 0.40 & -0.007788207(3) & -0.00799229975(7) & -0.0082942072(9) \\
 \hline
 $\pISO+1000$ & 0.00 & -0.000993413311181(0) & -0.000996016937701(0) & -0.000999595649400(0) \\
 $\pISO+1000$ & 0.10 & -0.000983179941(7) & -0.00098584074(1) & -0.000989496817(4) \\
 $\pISO+1000$ & 0.20 & -0.000953066056(2) & -0.000955719551(2) & -0.000959363369(3) \\
 $\pISO+1000$ & 0.30 & -0.000903093610(8) & -0.000905672227(9) & -0.000909211662(8) \\
 $\pISO+1000$ & 0.40 & -0.0008332886(4) & -0.000835722608(2) & -0.000839062961(5) \\
 \hline\hline
\end{tabular}
\caption{Numerical data for $\Delta U$ for a selection of orbits with  eccentricity $e$ varying from $0.0$ (circular) to $0.4$, and black hole spin $a=-0.9$ (retrograde), $a=0$ (Schwarzschild), or $a=0.9$ (prograde). The semi-latus rectum $p$ for each orbit has been chosen relative to the semi-latus rectum of the last stable orbit $\pISO$ at that eccentricity and spin. Parenthetical figures indicate the estimated error on the last displayed digit, with $(0)$ indicating that the error is at least an order of magnitude smaller than the last shown digit. Note that the data used for the location of $\pISO$ is accurate only to about 5 digits. For compactness, the table does not show the actual value of $p$ used for each orbit. This data is available as supplementary data.}\label{tab:ecckerr}
\end{table*}
Agreement with existing results in the relevant limits instils some confidence that our code is operating correctly. To further check our code we perform some consistency checks. 

We first check that the truncated sums have indeed converged to the expected level.
Fig. \ref{fig:lnconvplot} shows the convergence of the sum over $n$-modes for a sample orbit with spin $a=0.9$, semilatus rectum $p=3.32$, and eccentricity $e=0.2$. The low $l$-modes converge much faster than the high $l$-modes, but all $l$-modes eventually settle into an exponential convergence for large $n$. 

The characteristic two ``hump'' structure in Fig.  \ref{fig:lnconvplot} is the result of a qualitative difference in the convergence behaviour of the high and low $m$-modes. This behaviour is shown in Fig. \ref{fig:mnconvplot}, where we see a 3D representation of the $mn$-mode contributions to the $l$-modes. At low $n$, the $l$-mode is dominated by the contributions from the modes with small $m$. However, these contributions decay much more quickly than the large $m$ contributions. Consequently, at large values of $n$ the roles are reversed and the $l$-modes are dominated by $m=\abs{l}$ contributions.

A second apparent feature in Fig. \ref{fig:lnconvplot} is that some of the $n$-mode contributions to the (high) $l$-modes are much bigger (in magnitude) than the expected final value of the $l$-mode (as represented by the value of the regularization parameter $B$). This indicates that the sum over $n$-modes is very oscillatory and has a larger degree of cancellation. For this reason our convergence checks in section \ref{sec:truncation} are relative to $B$, aiming for a fixed accuracy.

Having established convergence of the truncated $n$-sums, we turn our attention to the $l$-modes. Fig. \ref{fig:lmodeplot} shows the values of the $l$-modes (again for the same reference orbit with spin $a=0.9$, semilatus rectum $p=3.32$, and eccentricity $e=0.2$) both for the outside ($\pI$) and inside ($\pH$) limits of $h_{uu}$. As expected, the values of the $l$-modes do not decay for large $l$. Instead the $l$-modes converge to some constant value. In section \ref{sec:reg} we argued  that the asymptotic behaviour of our radiation gauge should match the analytically known values for Lorenz gauge regularization parameters ($B$, $C$, and $D$). This can be checked numerically, by subtracting the analytical values of $B$ and $C$ from our numerical data. The results are also shown in Fig. \ref{fig:lmodeplot}. After subtraction of $B$, the $l$-modes decay as $l^{-2}$. This implies that our data is compatible with the value of both the $B$ and the $C=0$ terms. This agreement exists for both the outside ($\pI$) and inside ($\pH$) limits of $h_{uu}$. Similar agreement is found for other orbits. Besides providing a consistency check of our data, this is also the first numerical confirmation of the analytical calculation of the regularization parameters for orbits in Kerr for eccentric orbits.
\begin{figure}[tp]
\includegraphics[width=\columnwidth]{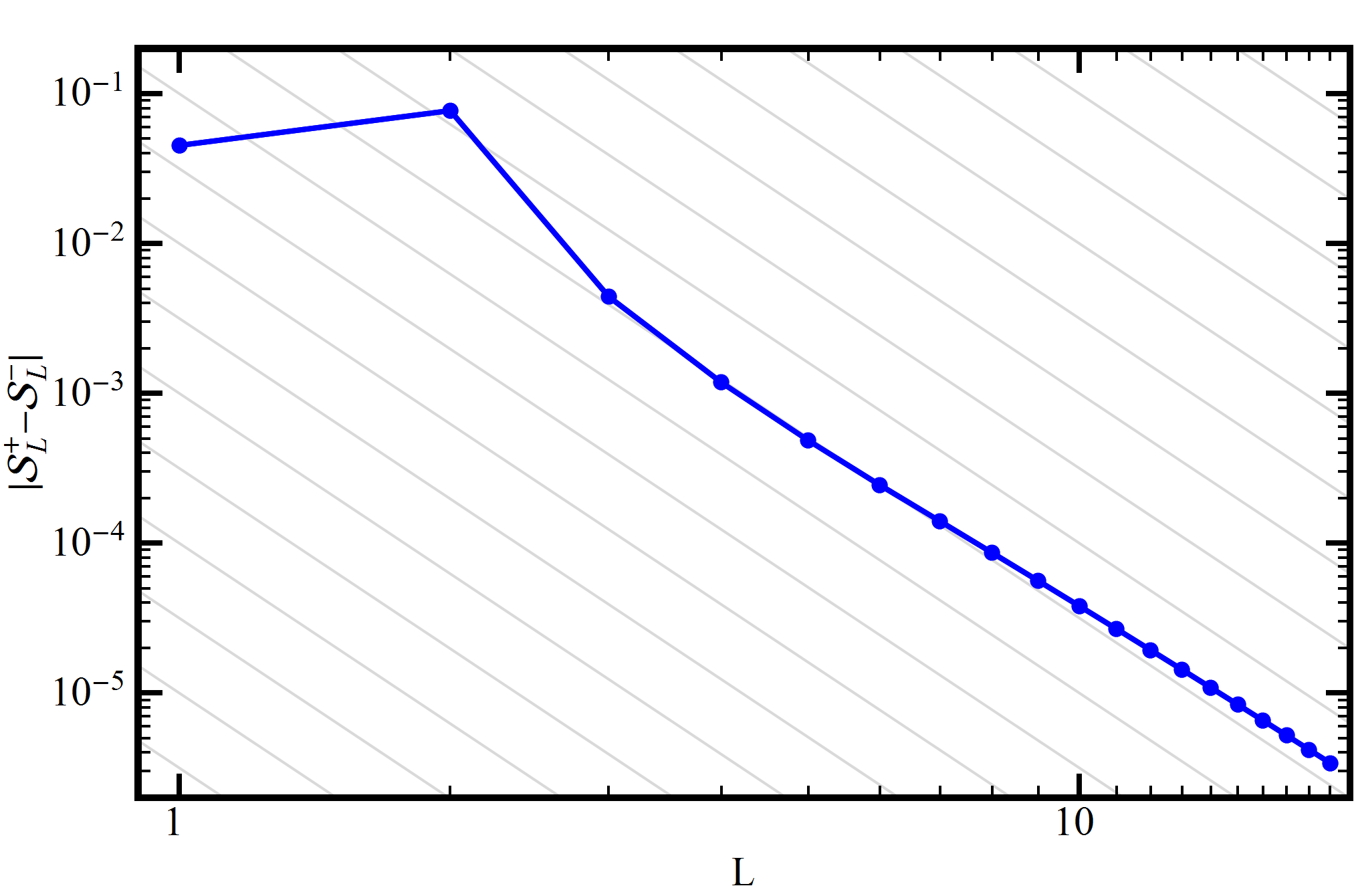}
\caption{The difference between the partial sums of the $l$-modes of $\avg{h_{uu}^\pIH}$ (excluding completion terms) of an orbit with spin $a=0.9$, semilatus rectum $p=3.32$, and eccentricity $e=0.2$ . As $L$ increases the difference decays as approximately $l^{-4}$ (diagonal gridlines).}\label{fig:inout}
\end{figure}

Not only do the $l$-modes of the reconstructed metric obtained from the interior and exterior regions have similar convergence behaviour, in fact they turn out to converge to the same value as can be seen in Fig. \ref{fig:inout}. At first sight, this is some what surprising since the exterior and interior metric perturbations are obtained in different radiation gauges. Since the gauge used to obtain the field in the interior vacuum region is irregular at infinity, we have no reason to expect $\avg{h_{uu}^{R,\pH}}$ to be the same as $\avg{h_{uu}^{R,\pI}}$ when we do not include the completion terms.
\begin{figure}
\includegraphics[width=\columnwidth]{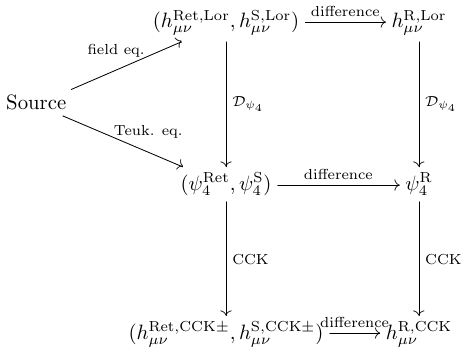}
\caption{Commutative diagram representing (an idealized version of) the CCK metric reconstrunction procedure followed in this paper. Commutativity of the diagram suggests that we should expect the inside anc outside limits of $h_{\mu\nu}^{\reg,\text{CCK}}$ to be the same. However, beware the caveat mentioned in the text.} \label{fig:cd}
\end{figure}

An intuitive understanding of why we would expect agreement can be gained from considering the diagram in Fig. \ref{fig:cd} representing an (idealized) version of the procedure used to obtain the regular metric perturbation. The vertical arrows in this diagram all represent linear operators, while the horizontal arrows are simple subtractions. Hence the diagram commutes. Following the left side of the diagram (roughly) corresponds to our procedure; Applying the CCK procedure to fields with a non-zero source leads to a gauge ambiguity between the values of the retarded and singular fields obtained from the interior and exterior vacuum regions. A similar ambiguity is thus expected in their difference, $h_{\mu\nu}^{\reg,\text{CCK}}$. However, if one follows the right-hand side of the diagram, then it follows from the fact that $h_{\mu\nu}^{\reg,\Lor}$ is a vacuum solution of the linearized Einstein equation that there is no ambiguity in applying the CCK procedure. Consequently, commutativity of the diagram implies that there is also no gauge ambiguity from other route.

It must however be noted that the above diagram is not exactly the procedure that we follow. In particular, we do not obtain the singular field by calculating the singular part of $\psi_4$ and applying the CCK procedure. Instead, the Lorenz gauge singular field is obtained as an asymptotic power series in $l$, and the radiation gauge version of this expansion is obtained by studying the effect of a local gauge transformation. It is not clear that this procedure will commute with the diagram. In particular, it is very well possible that this different route would still induce a gauge difference between pointwise values of regular metric perturbation in the interior and exterior. 

We conclude by giving values for $\Delta{U}$ for a selection of sample orbits around a spinning black hole in table \ref{tab:ecckerr}. On a single 16 processor node of our computing cluster, the computation time for these orbits ranges from about 2 minutes for the circular orbits to just over 5 and a half hours for the $0.4$ eccentricity orbits.

\section{Discussion and Conclusions}\label{sec:discussion}
In this paper we have described the first ever numerical computation of the gravitational perturbations produced by a small mass $m$ orbiting a Kerr black hole on an equatorial eccentric orbit to first order in $m$. This was achieved by applying the CCK metric reconstruction to a perturbed Weyl scalar calculated using the semi-analytical MST formalism.

As a demonstration of our code's capabilities we calculated the (generalized) redshift gauge invariant, $U$. In the regimes that were previously explored in the literature---circular equatorial orbits in Kerr and eccentric orbits in Schwarzschild---our code matches the previously available results.

The results for eccentric orbits around a Kerr black hole are completely new. They have not been calculated numerically, nor are there (to our knowledge) any post-Newtonian expansions of the generalized redshift available to compare against. Comparison against the analytically known regularization parameters verifies that the metric perturbations calculated by our code have the correct singular structure. It also provides the first numerical confirmation of the analytical calculation of the mode-sum regularization parameters for gravitational perturbations produced by eccentric orbits in Kerr.

The semi-analytical nature of the MST formalism at the basis of our numerical computation means that very high accuracies can be achieved at a relatively low cost. In particular, the method is unaffected by the numerical problems caused by the near degeneracy of ingoing and outgoing modes at very low frequencies, which fundamentally limited the accuracy of the Lorenz gauge frequency domain code of \cite{Akcay:2015pza}.

The main limiting factor for the accuracy of our code is the limit on the number of $l$-modes that can be calculated. This can be traced back to our ability to calculate the so-called ``renormalized angular momentum'' parameter $\nu$ that is needed to evaluate the MST series. This involves numerically finding the root of an algebraic equation involving continued fractions. This is only possible if a good enough initial guess for $\nu$ is available. For low frequency modes, this guess is provided by an analytic low frequency expansion. However, no such expansion is available for higher frequencies. To enable calculations in the strong field regime, our code relies on interpolation of a table of known high frequency modes. At this stage, this table has been populated up to $l=30$, limiting our code to $l\leq 30$ modes. However, with not too much effort it should be possible to populate this table to any value needed lifting this limitation.

For high eccentricities a secondary limiting factor to our accuracy is the number of $n$-modes needed to achieve the required accuracy. This limitation is inherent to the frequency domain approach. It can be overcome in two principle ways: more computation time or better efficiency. The first is currently still an option with the most demanding orbits in table \ref{tab:ecckerr} completing in just over 5 hours on a single node of our cluster. On the other hand, our code has not been fully optimized yet. For example, our code has many internal parameters controlling the internal working precision of various components of the code. These are currently set to ``safe'' high values to ensure loss of internal precision is never an issue. Optimization of these values could probably reduce runtime by an order of magnitude. Once optimal values for the internal precision have been established further performance could be gained by porting the most CPU intensive parts of the code to fixed precision C code.
 
With relatively little effort, we were able to substantially improve on the numerical estimates of the post-Newtonian expansion of the generalized redshift invariant presented in \cite{Akcay:2015pza}. It should be possible to obtain much better results by following the strategy set out in \cite{Johnson-McDaniel:2015vva} to estimate the post-Newtonian coefficients for the redshift from circular orbits. There the calculations were performed in the extreme weak field (with radii $\sim 10^{20}M$), where it is much easier to distinguish the $\log p$ terms in the expansion. In the extreme weak field very good estimates for the renormalized angular momentum exist, allowing the MST method to solve the field equations to essentially arbitrary precision, thousands of digits if necessary. With some tuning our new code should also work in that regime, and it should be possible to get extremely accurate results for the generalized redshift in the extreme weak field, both for Schwarzschild and Kerr black holes.
 
The method presented in this paper can in principle be applied to any quantity constructed from the metric and its derivatives. In particular, it should be possible to obtain the full self-force correction to the orbital motion (which requires covariant derivatives of the metric).  The first (and most CPU intensive) stage of the calculation to obtain the Hertz potential would remain identical. For the new quantities we would need the relevant counter parts of eq. \eqref{eq:huu4} and table \ref{tab:huuc}. However, the general procedure for obtaining them should be identical to the procedure set out in Sec. \ref{sec:MR}. The occurrence of higher derivatives of the metric simply requires the indices $i$ and $s$ to have a wider range. In addition, quantities involving derivatives of the metric will be more singular, and may therefore require a smoother choice of the extension in \eqref{eq:bext}, and final expression will contain a sum over several $\YB{j}{m}{l_1}{l_2}$. However, since the bulk of computational time is spent obtaining the Hertz potential, performance should be very similar to the implementation for $h_{uu}$.

In principle, it should also be possible to apply similar methodology to obtain the gravitational perturbations produced by a mass on a generic---eccentric and inclined---orbit around a Kerr black hole. Our solver for the Teukolsky equation \cite{Meent:2015a} can already handle such orbits. Moreover, the general techniques used here (extended homogeneous solutions, mode-by-mode CCK reconstruction) are equally applicable to the generic case. Obviously, the results in sec \ref{sec:comp} and \ref{sec:reg} would have to be generalized to the generic case. However, we see no reason to believe that there are any fundamental obstructions (other than complexity) to doing so.

For generic orbits, computational costs are expected to become an issue due to the number of modes that need to be calculated in the frequency domain. Some savings can however be attained by first considering eccentric inclined orbits exhibiting a (low order) $r\theta$-resonance \cite{Grossman:2011im}. Access to the full self-force on $r\theta$-resonant orbits should further allow verification for approximate results for the effect of such resonances using orbit averaged fluxes \cite{Flanagan:2012kg,Meent:2013}. It would also provide valuable input for the open question whether motion under the conservative self-force in Kerr is integrable \cite{Vines:2015efa}.

\section*{Acknowledgements}
This work has benefited from many useful discussions with Leor Barack and Adam Pound. We also wish to thank Sarp Akcay for providing numerical data from their code used in early validation of our own code. MvdM was supported by NWO Rubicon grant 680-50-1203. MvdM and AGS were supported by the European Research Council under the European Union's Seventh Framework Programme (FP7/2007-2013)/ERC grant agreement no. 304978. The numerical results in this paper were obtained using the IRIDIS High Performance Computing Facility at the University of Southampton.

\raggedright
\bibliography{journalshortnames,huukerr}

\end{document}

%% file: huutab.tex
\begin{tabular}{cc|l}
\hline\hline
 $i$ & $s$ & 
 $\MC_{is}(m,n,r_0)$\\
 \hline
 $0$ & $0$ & 
 $
 	\begin{aligned}[t]
		r_0^2u^1 u^1
	\end{aligned}
$ \\
 $0$ & $1$ & 
 $
	\begin{aligned}[t]
		-2a r_0(r_0 \omega_{mn} +i)u^1 u^1
		-\sqrt{2}\hh{i r_0 K_{mn} +2 (r_0^2- a^2)}u^1 u^3
	\end{aligned}
	$ \\
 $0$ & $2$ & 
 $
	\begin{aligned}[t]
		&a^2 r_0  \omega_{mn}  (r_0 \omega_{mn} +2 i) u^1 u^1
		+ \sqrt{2}a  \omega_{mn} \hh{i r_0 K_{mn}+2 (r_0^2- a^2)}u^1 u^3\\
		&+ \Bh{ \Delta_0  \bh{\frac{4- i K_{mn}}{r_0}+ i r_0 \omega_{mn} -2}-\frac{1}{2} \bh{K_{mn}-2 i (r_0-1)} \bh{K_{mn}-4 i (r_0-1)}} u^3 u^3
	\end{aligned}
	$ \\
 $1$ & $1$ & 
 $
 	\begin{aligned}[t]
		-\sqrt{2}\Delta_0  r_0 u^1 u^3
	\end{aligned}
	$ \\
 $1$ & $2$ &
 $
	\begin{aligned}[t]
		\sqrt{2}a \Delta_0  r_0  \omega_{mn} u^1 u^3
		+\Bh{-\frac{\Delta_0 ^2}{ r_0}+\Delta_0 \hh{i K_{mn}+4 (r_0-1)}}u^3 u^3 
	\end{aligned}
	$ \\
 $2$ & $2$ &
 $
 	\begin{aligned}[t]
		\frac{\Delta_0 ^2 }{2}u^3 u^3
	\end{aligned}
	$ \\
\hline\hline
\end{tabular}

%% file: circkerrtab.tex
\begin{tabular}{r|......}
\hline\hline
  r & \multicolumn{1}{c}{$a=-0.9$}	&	\multicolumn{1}{c}{$a=-0.7$}	&	\multicolumn{1}{c}{$a=-0.5$}	& \multicolumn{1}{c}{$a=0.5$}	& \multicolumn{1}{c}{$a=0.7$}	& \multicolumn{1}{c}{$a=0.9$} \\
\hline
  4 & \multicolumn{1}{c}{---} & \multicolumn{1}{c}{---} & \multicolumn{1}{c}{---} & \multicolumn{1}{c}{---} & -0.3934469(1) & -0.325705(1) \\
 \text{} & \text{} & \text{} & \text{} & \text{} & -0.3934452987(3) & -0.325704499(2) \\
 5 & \multicolumn{1}{c}{---} & \multicolumn{1}{c}{---} & \multicolumn{1}{c}{---} & -0.3135068(1) & -0.2776815(1) & -0.25072625(8) \\
 \text{} & \text{} & \text{} & \text{} & -0.3135069374 & -0.2776815518(1) & -0.2507261858(1) \\
 6 & \multicolumn{1}{c}{---} & \multicolumn{1}{c}{---} & \multicolumn{1}{c}{---} & -0.23426933(1) & -0.21718074(2) & -0.20325499(2) \\
 \text{} & \text{} & \text{} & \text{} & -0.2342693592(1) & -0.2171807422 & -0.2032550190 \\
 7 & \multicolumn{1}{c}{---} & \multicolumn{1}{c}{---} & \multicolumn{1}{c}{---} & -0.188583050(1) & -0.1788453416 & -0.170547666(3) \\
 \text{} & \text{} & \text{} & \text{} & -0.1885830414 & -0.1788453294 & -0.1705476660 \\
 8 & \multicolumn{1}{c}{---} & \multicolumn{1}{c}{---} & -0.204384597(2) & -0.158300271(2) & -0.152122681(2) & -0.146708039(2) \\
 \text{} & \text{} & \text{} & -0.2043845850 & -0.1583002707 & -0.1521226786 & -0.1467080374 \\
 10 & -0.151451799(1) & -0.1456824115(2) & -0.1404058589(3) & -0.1201650349(3) & -0.1171389690(2) & -0.1143966838 \\
 \text{} & -0.1514517993 & -0.1456824108 & -0.1404058593 & -0.1201650350 & -0.1171389696 & -0.1143966840 \\
 15 & -0.0833104689 & -0.0819468044 & -0.0806554028 & -0.0751903937 & -0.0742790117 & -0.0734230600 \\
 \text{} & -0.0833104689(1) & -0.0819468043(7) & -0.080655402(6) & -0.07519039(3) & -0.074279012(3) & -0.0734230600(6) \\
 20 & -0.0581474333 & -0.0575935465 & -0.0570620298 & -0.0547222331 & -0.0543144710 & -0.0539257152 \\
 \text{} & -0.0581474333 & -0.0575935465 & -0.0570620298 & -0.0547222331 & -0.0543144710 & -0.0539257152 \\
 30 & -0.0365055425 & -0.0363353216 & -0.0361700729 & -0.0354163457 & -0.0352797020 & -0.0351476286 \\
 \text{} & -0.0365055425 & -0.0363353216 & -0.0361700729 & -0.0354163457 & -0.0352797020 & -0.0351476286 \\
 50 & -0.0210263381 & -0.0209844570 & -0.0209434412 & -0.0207511788 & -0.0207152577 & -0.0206801699 \\
 \text{} & -0.0210263381 & -0.0209844570 & -0.0209434412 & -0.0207511788 & -0.0207152577 & -0.0206801699 \\
 70 & -0.0147844703 & -0.0147673393 & -0.0147504968 & -0.0146705787 & -0.0146554476 & -0.0146405986 \\
 \text{} & -0.0147844703 & -0.0147673393 & -0.0147504968 & -0.0146705787 & -0.0146554476 & -0.0146405986 \\
 100 & -0.0102349197 & -0.0102281718 & -0.0102215165 & -0.0101896244 & -0.0101835216 & -0.0101775103 \\
 \text{} & -0.0102349197 & -0.0102281718 & -0.0102215166 & -0.0101896245 & -0.0101835217 & -0.0101775104 \\
\hline\hline
\end{tabular}